\documentclass[twoside,english, iop]{emulateapj}
\pdfoutput=1

\makeatletter
\usepackage{amsmath}
\makeatother

\usepackage{babel}
\begin{document}

\title{APPLICATION OF GAS DYNAMICAL FRICTION FOR PLANETESIMALS:\\
 I. EVOLUTION OF SINGLE PLANETESIMALS}

\author{Evgeni Grishin \& Hagai B. Perets}

\affil{Physics Department, Technion - Israel Institute of Technology, Haifa,
Israel 3200003}
\begin{abstract}
The growth of small
planetesimals into large planetary embryos occurs
much before the dispersal of the gas from the protoplanetary disk.
The planetesimal - gaseous-disk interactions give rise to migration
and orbital evolution of the planetesimals/planets. Small planetesimals
are dominated by aerodynamic gas drag. Large protoplanets, $m\sim0.1M_{\oplus}$,
are dominated by type I migration \emph{differential} torque. There
is an additional mass range, $m\sim10^{21}-10^{25}g$ of \emph{intermediate
mass} planetesimals (IMPs), where gravitational interactions
with the disk dominate over aerodynamic gas drag, but for which such
interactions were typically neglected. Here we model these interactions
using the \emph{gas dynamical friction} (GDF) approach,
previously used to study the disk-planet interactions at
the planetary mass range. We find the critical size where GDF
dominates over gas drag, and then study the implications of GDF on
single IMPs. We find that planetesimals with small inclinations rapidly
become co-planar. Eccentric orbits circularize within a few Myrs,
provided the the planetesimal mass is large, $m\gtrsim10^{23}g$ and
that the initial eccentricity is low, $e\lesssim0.1$. Planetesimals
of higher masses, $m\sim10^{24}-10^{25}g$ inspiral on a time-scale
of a few Myrs, leading to \emph{an embryonic migration} to the inner
disk. This can lead to an over-abundance of rocky material (in the
form of IMPs) in the inner protoplanetary disk ($<1$AU) and induce
rapid planetary growth. This can explain the origin of super-Earth
planets. In addition, GDF damps the
velocities of IMPs, thereby cooling the planetesimal disk and affecting
its collisional evolution through quenching the effects of viscous
stirring by the large bodies. 
\end{abstract}

\section{INTRODUCTION}

Planets form in protoplanetary disks around young stars. Once km sized
planetesimals have been formed, their evolution is determined by three
basic dynamical processes: Viscous stirring, dynamical friction and
coagulation or disruption through collisions (see \citealp{2004ARAA..42..549G}
and references therein for details). These dynamical processes do
not include the effects from planetesimal-gas interaction in the disk,
which can be important during the early stages of planet formation
when gas is abundant (the first few Myr, with possible suggestions
for longer time-scales, \citealp{2014ApJ...793L..34P}).

During the evolutionary phases of planet formation, small dust grains
successively grow to large planetary embryos. For small size planetesimals,
aerodynamic gas drag is the dominant effect of the gas. It maintains
low relative velocities and keeps the orbits circular and co-planar.
Hence, the planetesimal disk is expected to be thin \citep{2002Icar..155..436O}.
It may also assist in the coagulation and merger of small bodies \citep{ormel+10,per+11}.
Large planetary embryos ($m\ge0.1M_{\oplus})$ can migrate due to
the interaction with the gaseous disk (see e.g. \citealp{2006RPPh...69..119P}
for a review and references therein.)

Gas-planetesimal interactions are therefore important both for low
mass planetesimals and large earth-sized or larger embryos. However,
there exists an\emph{ intermediate planetesimal} mass (IMP) range,
of the order of $m\sim10^{21}-10^{25}g$ in which gas-planetesimal
interactions are typically neglected, since such planetesimals are
large enough to be fully decoupled from the gas, and aerodynamic gas
drag is too weak to affect their gravitational dynamics, while they
are not sufficiently massive to exert significant torque on the ambient
gas and change its global properties \citep{1984Icar...60...29H,1999Icar..139..350T}.
Nevertheless, in this mass range, gravitational interactions with
the disk are not negligible, and the disk-planetesimal interactions
can be modeled through \emph{gas dynamical friction} (GDF), which
dominates over aerodynamic gas drag. Here we focus on the effects
of GDF on IMP and consider its role in their dynamical evolution.

\citet{1999ApJ...513..252O} showed that dynamical friction in gaseous
medium exerts a force even at low subsonic velocities. GDF is important
for various astrophysical systems: migration of globular clusters
in galactic gaseous haloes, in-spiral of binary stars in the companion's
envelope and merger and in-spiral of binary black holes (e.g. \citealt{2010MNRAS.402.1758S,2004ApJ...607..765E,2004cbhg.sympE..13E,2011ApJ...726...28B}).
Moreover, it was estimated that aerodynamic drag is comparable to
GDF for bodies as large as a few hundred km in diameter. Direct
applications of GDF on planetary systems with sufficiently large eccentricities
and inclinations have been recently studied \citep{2013MNRAS.428..658T,2013ApJ...762...21C,2012MNRAS.422.3611R,2011ApJ...737...37M},
however they deal with masses of fully formed planets, above the IMP
range. Various generalizations of GDF are found in the literature;
e.g. numerical modeling, circular orbits in the non-linear regime
and others (\citealt{2001MNRAS.322...67S,2007ApJ...665..432K,2008ApJ...679L..33K,2009ApJ...703.1278K,2010ApJ...725.1069K}),
and more recently \citet{2011ApJ...737...37M} obtained formulae for
GDF in 2D slab geometry, directly applicable to protoplanetary disks.

Motivated by recent developments both in GDF theory and its application
to planetary systems, we explore the implications of GDF on single
IMP. The implications on binary planetesimals will be discussed in
a subsequent paper (Grishin \& Perets, in prep.). To do so we review
and compare the effects of aerodynamic gas drag (Section \ref{sub:AERODYNAMIC-DRAG})
and GDF (Section \ref{sub:GAS-DYNAMICAL-FRICTION}) and derive the
typical size of planetesimals for which GDF dominates over gas drag
(Section \ref{sub:COMPARISON}). We then calculate the time-scales
for variation of orbital elements using analytical arguments, and study
GDF effects numerically (Section \ref{sec:GDF-EFFECTS}). Finally,
we discuss (Section \ref{sec:Discussion}) the implications of such
processes for the evolution planetesimals.

\section{GAS PLANETESIMAL INTERACTION}

\label{sec:GPI}

The standard models of evolution of planetesimal disks account for
various dynamical processes, including both physical processes due
to planetesimal interactions such as viscous stirring, dynamical friction
and physical collisions followed by coagulation, as well
as gas-planetesimal interactions through aerodynamic gas-drag on
small planetesimals, and planetary migration through disk torques
acting on large planetary embryos and planets.

Although gas-planetesimal interaction has been studied in the context
of small planetesimals, massive planetesimals have been largely ignored
as they are decoupled from the gas, and their size and velocity dispersion
is dominated by gravitational interactions. In the following, we revisit
this argument by carefully examining processes caused due to the presence
of gas and discuss their possible implications. We first review the
basic properties of aerodynamic gas drag and GDF respectively, and
then compare both forces and find the lower limit of the planetesimal
size for which GDF becomes dominant compared with aerodynamic gas
drag. The lower limit is found to be roughly in the order of a few
hundred km, compatible with \citet{1999ApJ...513..252O}'s estimation,
and depends on the position in the disk and its gas density.

\subsection{Aerodynamic gas drag}

\label{sub:AERODYNAMIC-DRAG}

The general drag force imposed on a planetesimal of radius $R$ and
relative velocity $v_{rel}$ moving through a gaseous medium of density
$\rho_{g}$ and speed of sound $c_{s}$ depends on $R/\lambda$, where
$\lambda=1/n\Gamma$, is the mean free path of the gas, $n$ is the
number density of the gas, and $\Gamma$ is the cross section of gas-gas
collisions.

When $R\lesssim\lambda$, Epstein regime applies where individual
scattering is considered. The drag force is 
\begin{equation}
\boldsymbol{F}_{D}=-\frac{4}{3}\pi R^{2}\rho_{g}\bar{v}_{th}\boldsymbol{v}_{rel}\label{eq:epstein}
\end{equation}
Where $\bar{v}_{th}=(8/\pi)^{1/2}c_{s}$ is the mean thermal velocity
(for Maxwellian distribution).

For $R\gtrsim\lambda$, the gas must be modeled as a fluid. Here,
the drag force depends also on the Reynolds number $\mathcal{R}e=2Rv_{rel}/\nu_{m}$,
where $\nu_{m}=(1/2)\bar{v}_{th}\lambda$ is the molecular viscosity
of the gas. For high Reynolds numbers ($\mathcal{R}e\gtrsim800$)
the gas exerts a ram pressure force, while for lower Reynolds numbers
Stokes drag is more applicable. The drag force is 
\begin{equation}
\boldsymbol{F}_{D}=-\frac{1}{2}C_{D}\pi\rho_{g}v_{rel}^{2}\hat{\boldsymbol{v}}_{rel}\label{eq:drag}
\end{equation}
where $C_{D}$ is the drag coefficient and $\hat{\boldsymbol{v}}_{rel}$
is the unit vector in the direction of relative velocity. Generally
$C_{D}$ depends on the geometry of the object, but for spherical
objects it depends only on Reynolds number, i.e. $C_{D}=C_{D}(\mathcal{R}e)$.
An empirical formula can be used for $C_{D}(\mathcal{R}e)$,
fitted for the range $\log_{10}\mathcal{R}e\in[-3,5]$ by \citet{2003JEE...129.222}.
\begin{equation}
C_{D}(\mathcal{R}e)=\frac{24}{\mathcal{R}e}(1+0.27\mathcal{R}e)^{0.43}+0.47[1-\exp(-0.04\mathcal{R}e^{0.38})]\label{eq:cd}
\end{equation}
In the large Reynolds number limit $C_{D}$ is constant, while in
the low Reynolds number limit $(\mathcal{R}e<1)$, $C_{D}\sim\mathcal{R}e^{-1}$.
From consistency of Eq. (\ref{eq:epstein}) and Eq. (\ref{eq:drag}),
the transition between Epstein and Stokes drag regimes is $R\approx(9/4)\lambda$,
and $C_{D}=24/\mathcal{R}e$ for $\mathcal{R}e<1$.

\subsection{Gas dynamical friction}

\label{sub:GAS-DYNAMICAL-FRICTION}

Consider a perturber with mass $m_{p}$ moving on a straight line
with constant velocity $\boldsymbol{v}_{rel}$ in a uniform gaseous
medium with density $\rho_{g}$ and characteristic sound speed $c_{s}$.
The perturber generates a wake, which in turn affects the perturber.
Using linear perturbation theory, \citet{1999ApJ...513..252O} calculated
the gravitational drag force felt by the perturber. The gas dynamical
friction (GDF) force is given by 
\begin{equation}
\boldsymbol{F}_{GDF}=-\frac{4\pi G^{2}m_{p}^{2}\rho_{g}}{v_{rel}^{3}}\boldsymbol{v}_{rel}\mathcal{I}(\mathcal{M})\label{eq:gdf}
\end{equation}
where $\mathcal{M}\equiv v_{rel}/c_{s}$ is the Mach number, and $\mathcal{I}(\mathcal{M})$
is a dimensionless factor given by 
\begin{equation}
\mathcal{I}(\mathcal{M})=\left\{ \begin{array}{cc}
\frac{1}{2}\ln\left(\frac{1+\mathcal{M}}{1-\mathcal{M}}\right)-\mathcal{M} & \mathcal{M}<1\\
\frac{1}{2}\ln\left(1-\frac{1}{\mathcal{M}^{2}}\right)+\ln\left(\frac{v_{rel}t}{R_{min}}\right) & \begin{array}{cc}
\mathcal{M}>1\\
v_{rel}t>R_{min}
\end{array}
\end{array}\right.\label{eq:I of m}
\end{equation}

The force is non-vanishing in the subsonic regime, while in the supersonic
regime, a minimal radius $R_{min}$ is introduced to avoid divergence
of the gravitational potential (usually taken to be the physical size
of the perturber, or the accretion radius $Gm_{p}/v_{rel}^{2}$).
The exact value of $R_{min}$ is not well determined; but it can be
fitted through comparison of Eq. (\ref{eq:I of m}) with hydrodynamical
simulations, to find a best fitting value for $R_{min}$ \citep{1999ApJ...522L..35S}.
It is important to stress that Ostriker's original calculation was
done using a point mass perturber; GDF is a gravitational volume force,
essentially different from the aerodynamic drag, which is a surface
force, dependent on the geometry of the perturber.

For small Mach numbers $\mathcal{M}\ll1$, 
\begin{equation}
\mathcal{I}(\mathcal{M})=\mathcal{M}^{3}/3+O(\mathcal{M}^{5})\label{eq:smallmach}
\end{equation}

We note that the results of \citet{2007ApJ...665..432K} are qualitatively
similar to \citeauthor{1999ApJ...513..252O}'s straight line trajectory
(see fig. (8) of \citealp{2007ApJ...665..432K}). Comparison to \citet{2011ApJ...737...37M}'s
formulae for slab geometry and validity of both models is discussed
in Section \ref{sub:Caveats}.

\subsection{Comparison between gas dynamical friction and aerodynamic gas drag }

\label{sub:COMPARISON}

Aerodynamic drag is more dominant for small size planetesimals and
scales as $\sim R^{2}$.\footnote{For ram pressure regime $C_{D}$ is roughly constant, so this is indeed
the case. For different drag regimes $C_{D}$ weakly depends on $R$
via the Reynolds number. For wide range of Reynolds numbers, $F_{D}\sim R^{\alpha}$
where $1\le\alpha(\mathcal{R}e)\le2$. } GDF is negligible for small sizes, and scales as $\sim R^{6}$, since
$m\sim\rho_{m}R^{3},$ where $\rho_{m}$ is the material density.

It is therefore clear that there exists a unique value $R_{\star}(G,\rho_{m},v_{rel},\mathcal{R}e,\mathcal{M})$
for which the aerodynamic drag and GDF forces are equal. Moreover,
$R_{\star}$ is not dependent on $\rho_{g}$. The only independent
dimensional parameters are $G,\rho_{m}$ and $v_{rel}$. Dimensional
analysis shows that $R_{\star}$ scales as $R_{\star}\propto v_{rel}(G\rho_{m})^{-1/2}$,
where the proportion constant depends on the dimensionless numbers
$\mathcal{R}e,$ and $\mathcal{M}$. Comparing equations (\ref{eq:drag})
and (\ref{eq:gdf}) then yields the critical size 
\begin{eqnarray}
R_{\star} & = & 0.29\left[\frac{C_{D}(Re)}{\mathcal{I}(\mathcal{M})}\right]^{1/4}\frac{v_{rel}}{\sqrt{G\rho_{m}}}\label{eq:critical size}
\end{eqnarray}

The radial drift of the gas is slightly sub-Keplerian due to pressure
gradients. For circular orbits, the radial drift velocity of the gas
is $\boldsymbol{v}_{gas}=\boldsymbol{v}_{K}\sqrt{1-\eta},$ where
\begin{equation}
\eta=-\frac{\partial P/\partial\ln a}{2\rho_{g}v_{K}^{2}}\label{eq:eta}
\end{equation}
\citep{2010AREPS..38..493C}. For flat razor-thin disk, $\Sigma\propto a^{-1}$
, $P\propto a^{-3}$ we get $\eta=3H_{0}^{2}$, where $H_{0}=h/a\sim c_{s}/v_{K}$
is the aspect ratio of the disk \citep{2013apf..book.....A}.

Consider a planetesimal travelling in a circular orbit in a gaseous
disk. The relative velocity of the headwind is $\boldsymbol{v}_{rel}=\boldsymbol{v}_{K}-\boldsymbol{v}_{gas}\equiv K\boldsymbol{v}_{K}$.
It is convenient to define $\varepsilon=1-\sqrt{1-\eta}\approx\eta/2+\mathcal{O}(H_{0}^{4})$.
Thus, using (\ref{eq:eta}), $K=\varepsilon$. For typical ranges
of $H_{0}\in[0.01,0.05]$, the range of $\varepsilon$ is $\approx[10^{-4},4\cdot10^{-3}]$.

\begin{figure}[h]
\begin{centering}
\includegraphics[height=5.5cm]{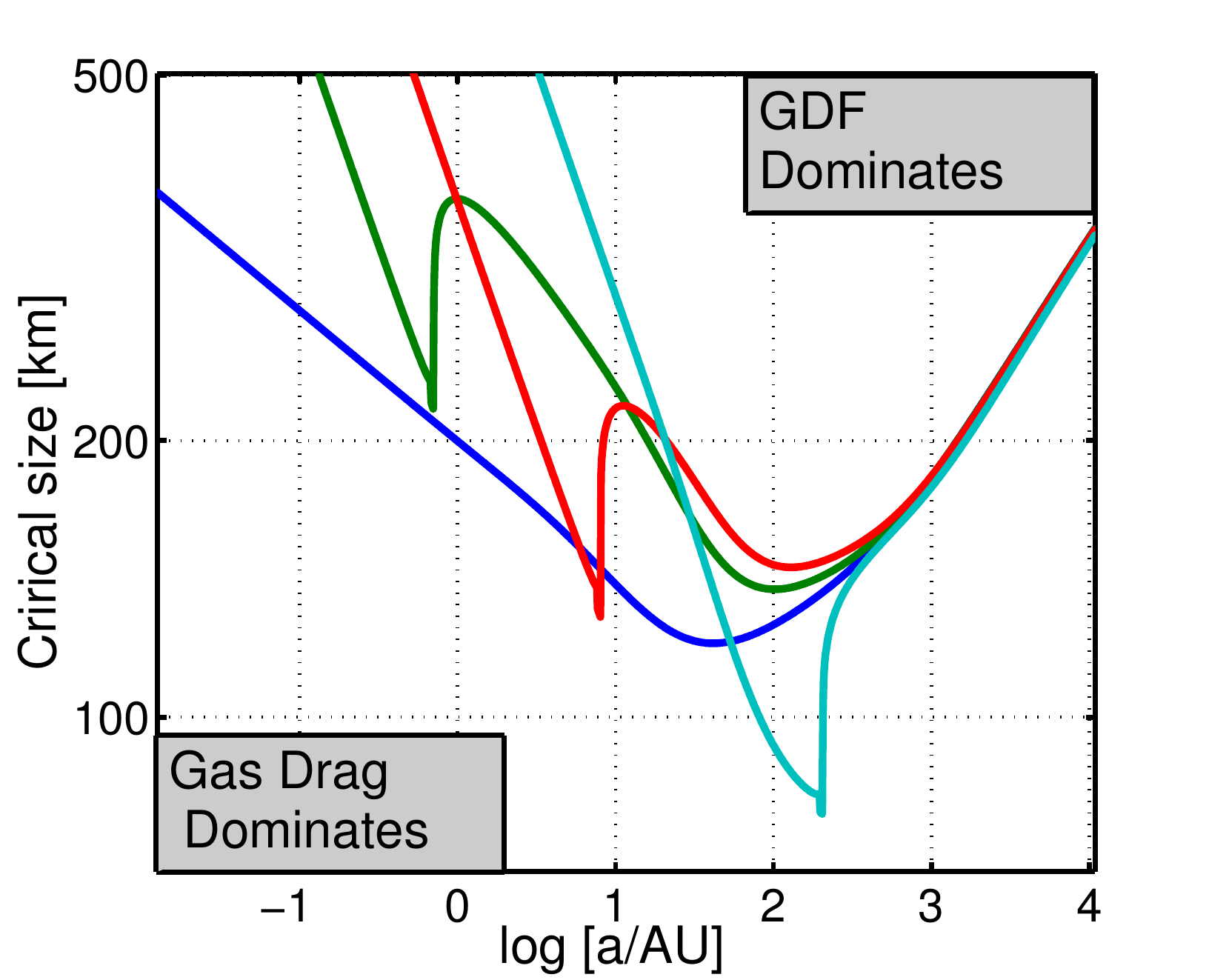}
\includegraphics[height=5.5cm]{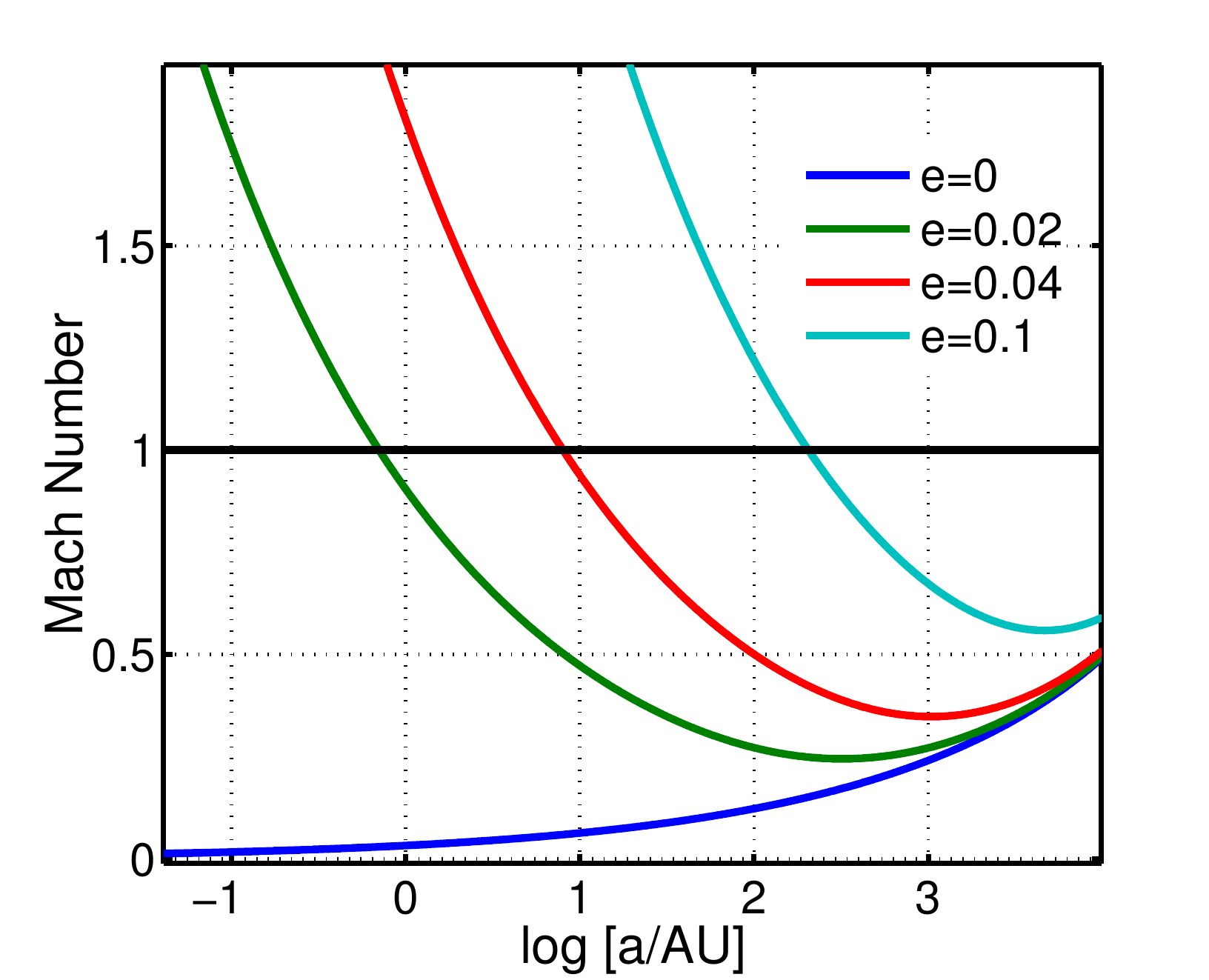} 
\par\end{centering}

\protect\protect\caption{\label{fig:1-1} Numerical solution for equation (\ref{eq:critical size})
for various eccentricities. Top: Critical size of planetesimal as
a function semi-major axis. Each curve corresponds to different orbital
eccentricity $e$, and indicates the critical size $R_{\star}(a)$
where both forces are equal. Aerodynamic drag dominates for smaller
radii, while GDF dominates for larger radii. Note log-log scale. Bottom:
Estimated Mach number as a function of semi-major axis for the same
orbits. Note the logarithmic scale.}
\end{figure}

For a concrete example, we consider a disk with similar parameters
as used by \citet{per+11}. The disk scaling is adapted from \citet{1997ApJ...490..368C}'s
simple flared disk model. Let us consider a specific power law scaling
of disk parameters, with respect to distance, $a$, to the central
star. Disk temperature is $T=120(a/AU)^{-3/7}K$. Molecular weight
is $\mu=2.3m_{H}$, where $m_{H}=1.66\cdot10^{-24}g$ is the hydrogen
atom mass. This leads to $c_{s}=(k_{B}T/\mu)^{1/2}=6.63\cdot10^{4}(a/AU)^{-3/14}cm\cdot s^{-1}$.
The disk aspect ratio is $H_{0}=h(a)/a=0.022(a/AU)^{2/7}$, where
$h(a)$ is the disk scale height.

For full evaluation of the relative velocities we will relax our assumption
of circular orbits and let the eccentricity, $e$, be a free parameter.
We show in appendix \ref{sec:Typical-eccentricity-for} that
the ratio of the relative and the Keplerian velocity $K=v_{rel}/{v_{K}}$
alters to $K\approx\sqrt{\varepsilon^{2}+e^{2}}+\mathcal{O}(e^{4})$.
A more complete treatment of the relative velocity between an eccentric
orbit and a planet is given in \citet{2011ApJ...737...37M}. The Reynolds
number is $Re=2Rv_{rel}/\nu_{m}$, where $R$ is the planetesimal
size, $\nu_{m}\sim c_{s}\lambda$ is the molecular viscosity, and
$\lambda$ is the mean free path. Note that although Eq. (\ref{eq:critical size})
appears simple, $C_{D}$ depends on $\mathcal{R}e$ via Eq. (\ref{eq:cd}),
which in turn depends on $R_{\star}$.

The top panel of Fig. \ref{fig:1-1} shows the numerical solution of Eq.
(\ref{eq:critical size}). We see that for low eccentricities, planetesimals
with $R\gtrsim500km$ are dominated by GDF for most regions of the
disk. We see that the derivative of the solution $R_{\star}(a,e)$
is discontinuous where the Mach number $\mathcal{M\approx}1$. It happens
approximately where $e=H_{0},$ the disk aspect ratio. The origin
of the discontinuity is the simplified discontinuous behavior of $\mathcal{I}(\mathcal{M})$
near $\mathcal{M\approx}1$ given in Eq. (\ref{eq:I of m}).

The bottom panel of Fig. \ref{fig:1-1} shows the dependence of the
orbit averaged Mach number on $a$. For $H_{0}\ll e$, $v_{rel}=ev_{K}$
and $\mathcal{M}=v_{rel}/c_{s}\propto a^{-2/7}$, i.e. as long as
eccentricity is high, the Mach number is a decreasing function of
$a$. For $H_{0}\gg e$ , $v_{rel}=\varepsilon v_{K}\approx H_{0}^{2}v_{K}$
and $\mathcal{M}\propto a^{2/7}$ so the Mach number is an increasing
function of $a$. The transition occurs when $H_{0}$ is a few times
the orbital eccentricity.

In the next sections we will investigate the effects of GDF on single
IMPs.

\section{ GDF EFFECTS ON THE EVOLUTION OF SINGLE PLANETESIMALS}

\label{sec:GDF-EFFECTS}

Current studies of gas planetesimal interactions have typically been
restricted to the effects of aerodynamic gas drag forces (\citealp{2010EAS....41..187Y,2004ARAA..42..549G}).
Here, we implement the effects of GDF for IMPs.

\subsection{Formulation of the problem}

\label{sub:formulation}

Consider a planetesimal of mass $m_{p}$ traveling with relative velocity
$v_{rel}$ with instantaneous orbital parameters $(a,e,I)$, where
$a$ is the semi-major axis, $e$ is the eccentricity and $I$ is
the inclination angle.

First, we compute the migration time-scale due to GDF of a circular
orbit, and then compare it to N-body simulations with external drag
force. For now, we set the inclination to zero, thus the problem has
two dimensions. Later on we consider inclined orbits and relax this
assumption.

Let us consider a typical gaseous disk with $M_{gas}/M_{\star}\sim H_{0}\sim0.01$
($M_{gas}/M_{\star}\sim0.1$ for most massive disks). Above this limit
self gravity is non-negligible and the disk becomes unstable (e.g.
\citealp[ and references therein]{2013apf..book.....A}). A typical
intermediate mass planetesimal has a diameter of $200-5000$ km, and
corresponding mass of $10^{21}-10^{25}g$, where the higher end can
already be considered to be a planetary embryo.

We use \citet{1999ApJ...513..252O}'s linear theory. Deviations from
linear theory and other models are discussed later on.

The problem of accretion of gas by a spherical body was first studied
by \citet{1952MNRAS.112..195B}. The effective radius of accretion
is Bondi radius $r_{B}=2Gm/c_{s}^{2}$. In our case, $r_{B}/r=8\pi G\rho_{m}r^{2}/3c_{s}^{2}=1.25(r/1000km)^{2}.$
Thus for masses lower than $10^{25}g$, accretion is negligible. For
a planetary embryo, we estimate the accretion rate by taking $50\%$
accretion efficiency with $\dot{m}\sim0.5\pi r^{2}\rho_{g}c_{s}\sim10^{12}g\cdot s^{-1}$.
It will take $\tau_{acc}\sim m/\dot{m}\sim1Myr$ for a planetesimal
to accrete its own mass. For simplicity we neglect accretion in our
simulations. We'll briefly discuss the implications of accretion in
the summary.

\subsection{Timescales for the variation of the orbital elements}

\label{sub:VARIATION--DUE}

Consider a distorting force $\mathbf{F}=F_{r}\hat{\boldsymbol{r}}+F_{\varphi}\hat{\boldsymbol{\varphi}}$.
The variation of orbital elements can be calculated analytically.
The change in the semi-major axis is \citep{1999ssd..book.....M}
\begin{equation}
m_{p}\frac{da}{dt}=2\frac{a^{3/2}}{\sqrt{GM_{\star}(1-e^{2})}}[F_{r}e\sin f+F_{\varphi}(1+e\cos f)]\label{eq:dadt}
\end{equation}
and 
\begin{equation}
m_{p}\frac{de}{dt}=\sqrt{\frac{a(1-e^{2})}{GM_{\star}}}[F_{r}\sin f+F_{\varphi}(\cos f+\cos E)]\label{eq:dedt}
\end{equation}

where a is the semi-major axis $e$ is the orbital eccentricity and $f$ and $E$ are the true and eccentric anomalies respectively. The relationship between the true and eccentric anomalies is $\tan(f/2)=\sqrt{(1+e)/(1-e)}\tan(E/2)$. For small eccentricies, both anomalies coincide.

The GDF force depends on the relative velocity, hence a crucial step
in calculating the interaction between gas and planetesimals is evaluating
the relative velocities. The general case is hard to estimate analytically
and requires averaging the relative velocity on orbital period (see
also \citealp{2011ApJ...737...37M}), but for small eccentricites (i.e. using  equations \ref{eq:gdf} and \ref{eq:smallmach} the components of the force are $F_{r}=F_{0} v_{rel,r}$ and $F_{\varphi}=F_{0} v_{rel,\varphi}$, where $F_{0}=4\pi G^{2} m_{p}^{2} \rho_{g} / 3 c_{s}^3$ and $v_{rel,r}$ and $v_{rel,\varphi}$ are given by eq's (\ref{eq:relr}) and (\ref{eq:relphi}). Neglecting the $(1-e^{2})^{1/2}$ term in the denominator and assuming $ \eta \ll 1 $, the relative velicities are $v_{rel,r}/v_{K} \approx e \sin f$ and $v_{rel,\varphi}/v_{K} \approx -\eta/2  - e\cos f $, which leads to a simplied formulae for Eqs. (\ref{eq:dadt}) and (\ref{eq:dedt})
\begin{equation}
\frac{\dot{a}}{a} = 2\frac{F_{0}}{m_{p}}\left[-e^{2}\cos(2f)-\frac{\eta}{2}+\left(1-\frac{\eta}{2}\right)e\cos f\right]
\end{equation}
and
\begin{equation}
\frac{\dot{e}}{e} = \frac{F_{0}}{em_{p}}\left[-e(\cos(2f)+\cos^{2}f)+\eta\cos f\right]
\end{equation}
averaging over one orbit leads us to $\langle \dot{a}/a \rangle = - \eta F_{0}/m_{p}$ and $\langle \dot{e}/e \rangle = -  F_{0}/2m_{p}$, where $ 2\pi \langle A \rangle \equiv  \intop_{0}^{2\pi} Adf$ .   The typical time-scale for in-spiral is then 
\begin{eqnarray}
\tau_{a} & =|\langle \frac{a}{\dot{a}} \rangle| & =\frac{3}{4\pi\eta}\frac{c_{s}^{3}}{G^{2}\rho_{g}m_{p}}\nonumber \\
 & \approx & 5.7\cdot10^{6}\left(\frac{\rho_{g}}{\rho_{0}}\right)^{-1}\left(\frac{m_{p}}{2\cdot10^{25}g}\right)^{-1}yr\label{eq:tcirc-1}
\end{eqnarray}

Where $\rho_{0}=3\cdot10^{-9}g\cdot cm^{-3}$. For $m=2\cdot10^{25}g$
, $\tau_{a}\sim5.7Myr$, which is comparable with the disk lifetime.
Smaller masses are less affected by GDF over the disk lifetime (see
also Section \ref{sub:pmass} ). Note that $\tau_{a}$ is inversely
proportional to $\rho_{g}.$ At larger separations, $\rho_{g}$ decreases
and GDF is less effective. At lower separations GDF stronger, and
the rate of in-spiral increases. Thus, $\tau_{a}$ is an upper limit.

For small eccentricity, the eccentricity damping timescale is much faster than the migration timescale (i.e. $\tau_{e}/\tau_{a} = 2\eta = 6H_{0}^{2}$ ). In addition, the eccentricity decays exponentially, i.e. $\dot{e} \propto e $. The latter is consistent with the analysis of eccentrcity waves raised by a planet studied by \citealp{2004ApJ...602..388T}. They find that the decay is exponential and that $\tau_{e} \approx H_{0}^{2} \tau_{a} $ .For larger eccentricities, i.e.  $e\gtrsim H_{0}$, the calculation is more complex. \citet{2000MNRAS.315..823P} find $\dot{e}\propto e^{-2}$
and
\begin{equation}
\tau_{e}/\tau_{a}\approx e^{2} \label{largee}
\end{equation} 
Both the exponential and the power
law decay are confirmed in numerical simulations of type I migration
\citep{2007AA...473..329C}. The connection between type I migration
and GDF is discussed in section \ref{typeI} 

Generally, we expect that the eccentricity to be damped faster than
semi-major axis.

The inclination decays rapidly compared with the other orbital elements.
Orbital inclination changes due to the $\hat{\boldsymbol{z}}$ component
of GDF. For an inclined orbit with inclination angle $I$, the normal
force is proportional to $F_{z}\propto v_{z}\sim Iv_{K}$. For a circular
orbit, it is much larger than $F_{\varphi}$, since the ambient gas
does not possess any significant $\hat{z}$ component for gas velocity.
The inclination decay time is then $mdI/dt=\sqrt{a/GM_{\star}}F_{z}\cos(\omega+f)$.
Most of the force is applied during the planetesimal passage through
the bulk of the disk, i.e. where $\cos(w+f)\sim1$. 
\begin{equation}
\tau_{I}/\tau_{a}=\frac{\dot{I}}{I}\frac{a}{\dot{a}}=2I\frac{F_{\varphi}}{F_{z}}\sim2I\frac{\varepsilon v_{K}}{Iv_{K}}=2\varepsilon\sim10^{-3}\label{eq:tinc}
\end{equation}
and we therefore expect that initially inclined orbits will be damped
on much shorter time-scales. For $m\sim2\cdot10^{25}g$, $\tau_{I}\sim8\cdot10^{3}yr$.
\citet{2012MNRAS.422.3611R} found that for highly inclined
orbits, $\tau_{I}/\tau_{a}\sim I\sin^{2}(I/2)/\sin I$ (see Eqs. 10
and 11 in \citealp{2012MNRAS.422.3611R}). In the limit of small inclination,
$I\lesssim H_{0}$, it reduces to $\tau_{I}/\tau_{a}\sim I^{2}\sim H_{0}^{2}$,
which is consistent with Eq. (\ref{eq:tinc}) since in section \ref{sub:COMPARISON}
we have shown that $\varepsilon\sim H_{0}^{2}$ . We see that generally
orbital eccentricity and inclination are damped faster than semi-major
axis, hence most of the orbits will in-spiral in time-scale comparable
to circular orbit given in Eq.(\ref{eq:tcirc-1}). For larger
inclinations the decay time-scale will increase.

Inclined orbits evolve in three dimensions, hence a vertical structure
of the gas density must be introduced. We assume a standard Gaussian
vertical structure of the form $\rho_{g}\sim\exp(-z^{2}/2h^{2})$
\citep{2010EAS....41..187Y}, where $h$ is the disk scale height.
For inclinations much larger than the disk scale height, we expect
$\tau_{I}$ to be be larger, since the planetesimal spends most of
its evolution above or below the disk bulk where gas density is low.

Now that we developed a basic analytic qualitative understanding of
the expected effects of GDF, we continue to a more detailed quantitative
study of these effects using numerical simulations.

\subsubsection{Numerical set up}

To study the effects of GDF we use an N-body integrator with a shared
but variable time step, using the Hermite 4th order integration scheme
following \citet{1995ApJ...443L..93H}. In order to include GDF effects
we add a fiducial GDF force that mimics Eq. (\ref{eq:gdf}). At each
step we calculate the additional external acceleration and jerk due
to GDF. Full description of the calculation can be found in the appendix
\ref{sec:Numerical set up}.

\subsection{Results}

\begin{figure}
\includegraphics[height=5.8333cm]{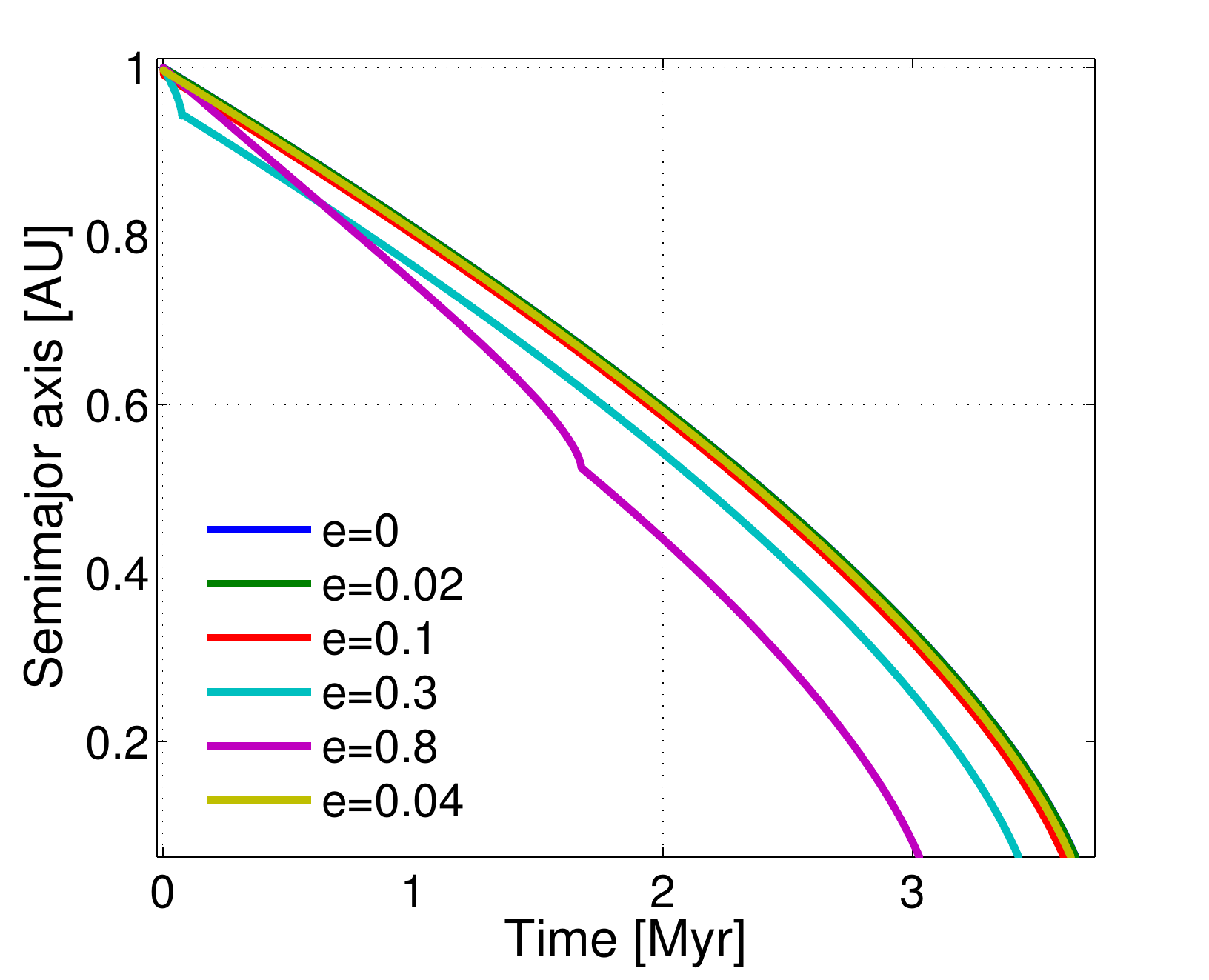}
\includegraphics[height=5.8333cm]{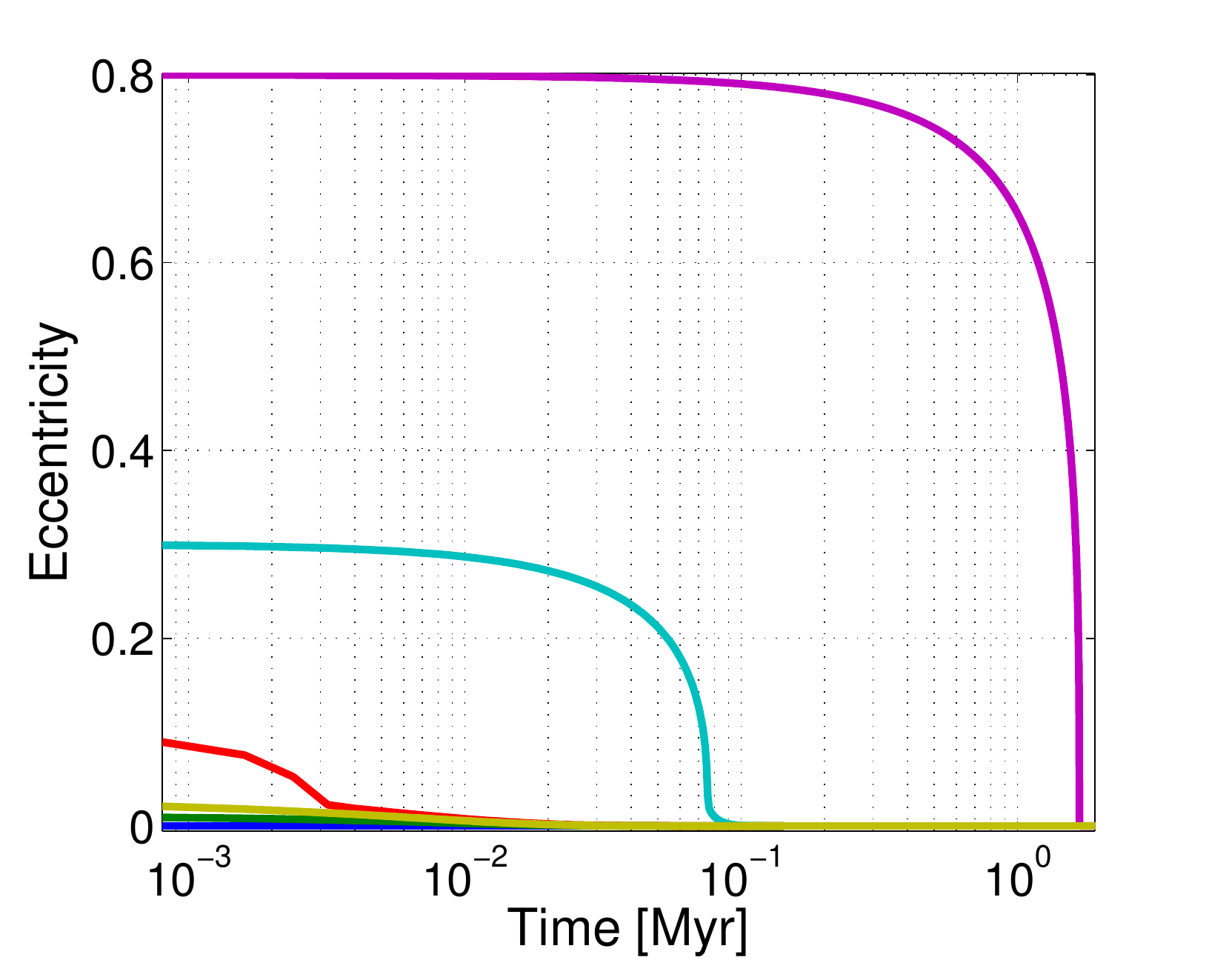}
\begin{raggedright} \includegraphics[height=5.75cm]{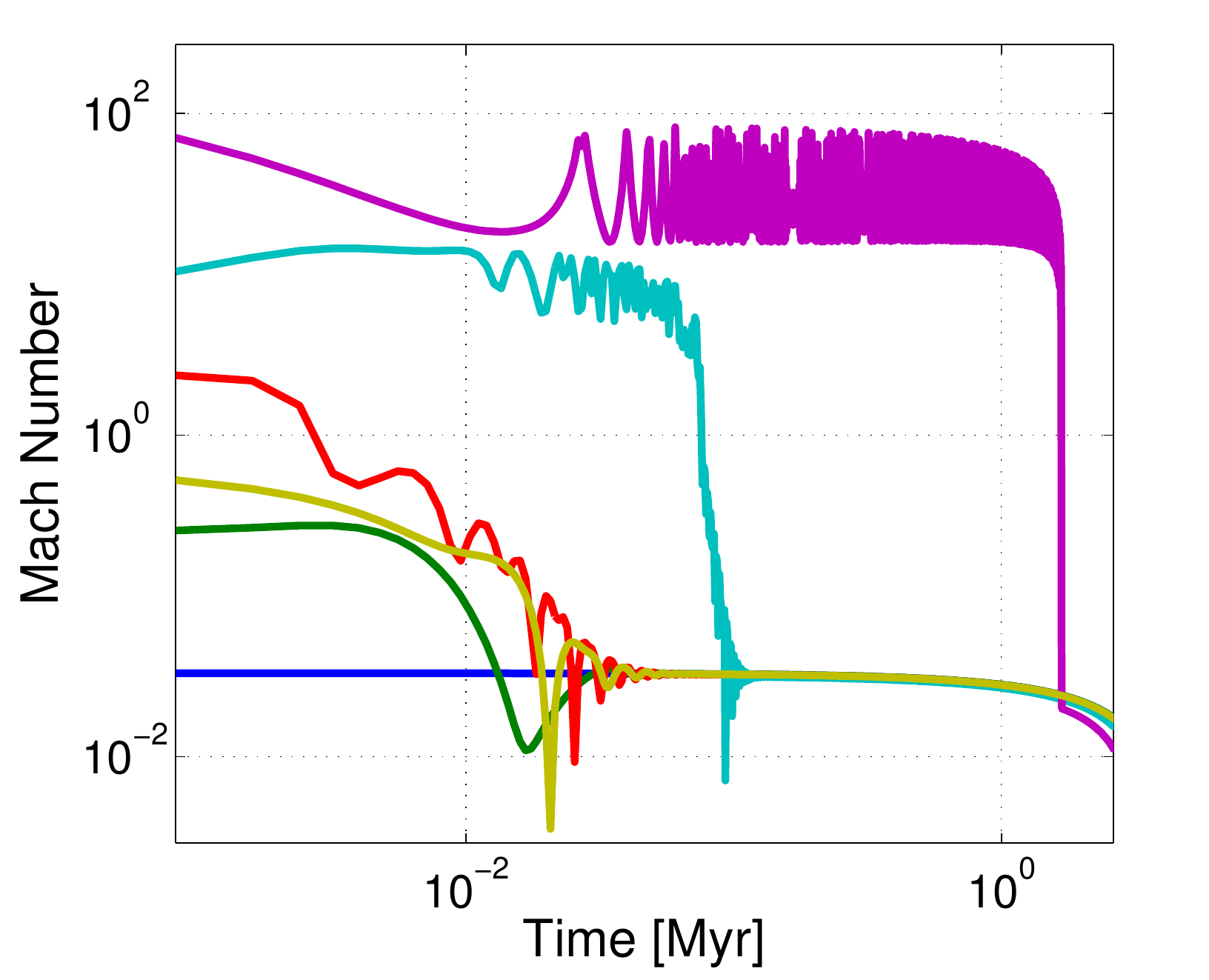} 

\end{raggedright}

\protect\protect\caption{\label{fig:Evolution-of-orbital}Evolution of orbital elements of
various orbits starting at $a=1AU$ and different eccentricities.
The mass of the planetesimal considered here is $2\cdot10^{25}g$.
Top: Evolution of the semi-major axis. Middle: evolution of the eccentricity.
Note logarithmic scale. Bottom: Evolution of the Mach number. Note
logarithmic scale. }
\end{figure}

In the following we present the results of a single planetesimal evolution
and effects of GDF on its evolution where various types of orbits
and planetesimal masses are considered.

\subsubsection{Eccentric orbits}

On the top panel of figure \ref{fig:Evolution-of-orbital} we see
that all the simulated orbits evolve in a similar manner. This is
mostly due to the rapid circularization of the orbits; as the orbits
circularize the relative gas-planetesimal velocities decrease, with
a corresponding decrease of the Mach number.

Eventually, their Mach number decreases to $\mathcal{M}\approx1$, at
which stage GDF is highly efficient, leading to a rapid loss of the
angular momentum. The larger the initial eccentricity is, the longer
it takes the orbit to reach the trans-sonic limit. Generally, the
eccentricity damping time-scale given by Eq. (\ref{largee}) 
is compatible with simulations. Actually, Eq. (\ref{largee}) overestimates
$\tau_{e}$. More rigorous derivation could determine $\tau_{e}$
more accurately.

We see that in most cases the planetesimals migrate within $3$ Myrs,
compatible with the estimation of $\tau_{a}$ in (\ref{eq:tcirc-1}).
The greater the initial eccentricity of the orbit is, the faster it
spirals in. This is due to the enhanced gas density $(\rho_{g}\propto a^{-16/7})$
in the apastron. For smaller planetesimals, e.g. of mass $\sim10^{23}g$,
$\tau_{a}$ increases to $\sim200Myr$, much more than typical disk
lifetimes. Note that given the short circularization time of eccentric
orbits the time-scale for the evolution of the semi-major axis is
almost independent of the initial eccentricity, and orbits with initial
eccentricities of $e\lesssim0.3$ are essentially circular through
most of their orbital evolution.

Orbits with initial eccentricity of $e=0.1$ circularize to $\sim0.02$
over $10^{4}$ years, and are fully circularized after $10^{5}$ years.
For orbits of initial eccentricity $e=0.3,$ the circularization time
extend to $\sim0.1$Myrs. For smaller masses, the lowest mass for
which the $\tau_{e}$ is comparable to typical disk lifetimes is $\sim10^{23}g$,
which corresponds to $500$km planetesimals. Less massive planetesimals
hardly evolve within the disk lifetime.

\subsubsection{Mildly inclined orbits}

\begin{figure*}
\begin{centering}
\includegraphics[height=5cm]{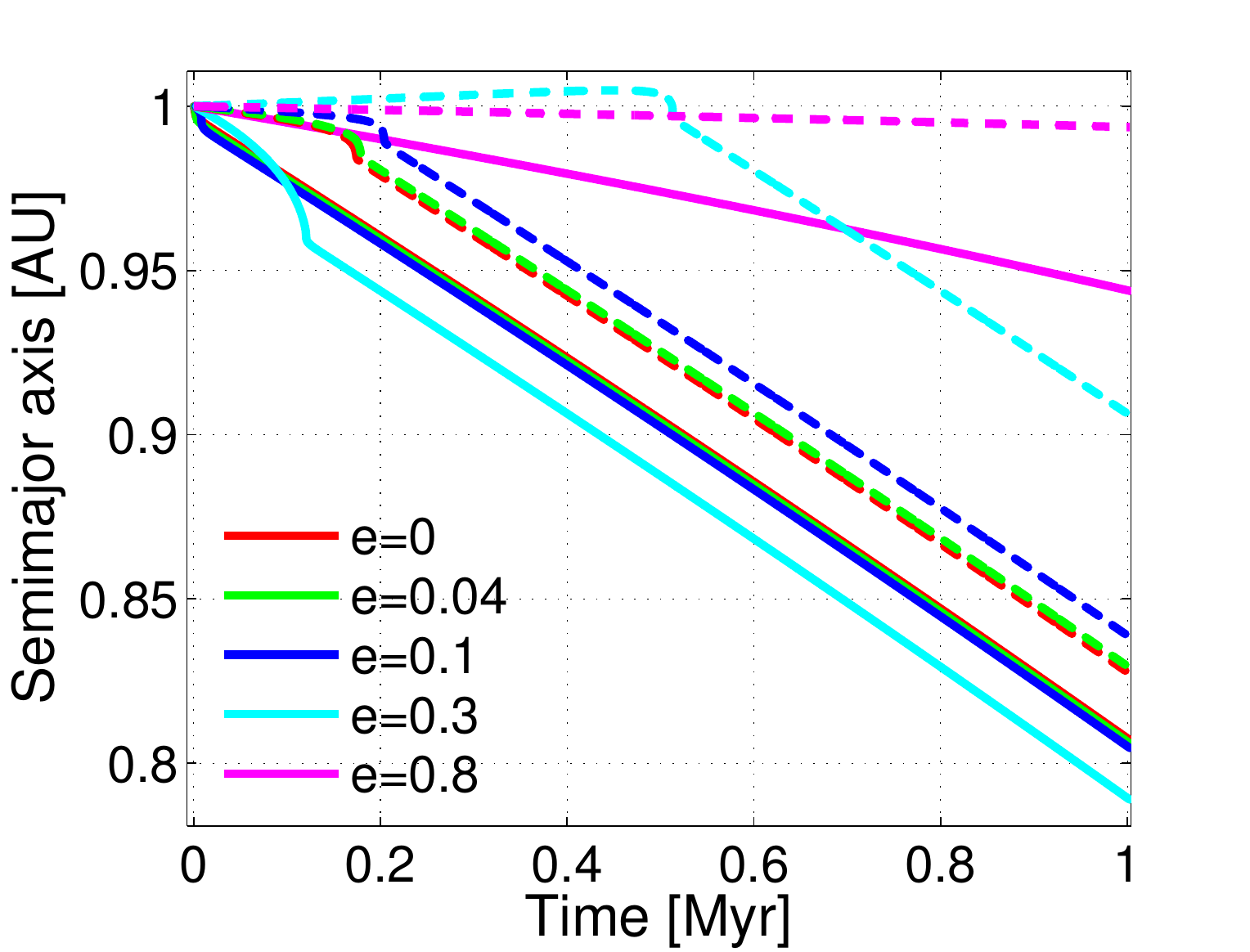}\includegraphics[height=5cm]{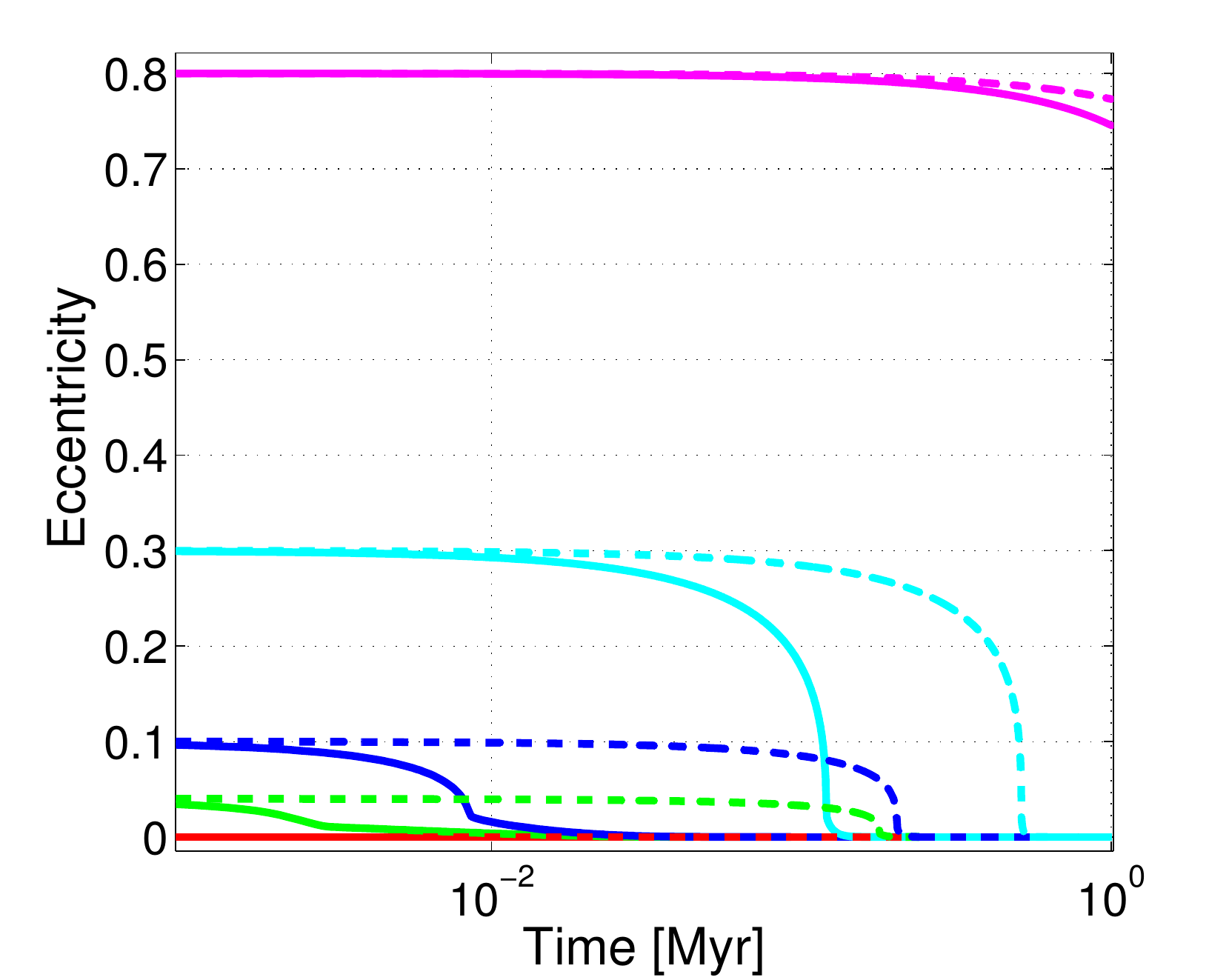} 
\par\end{centering}

\begin{centering}
\includegraphics[height=5cm]{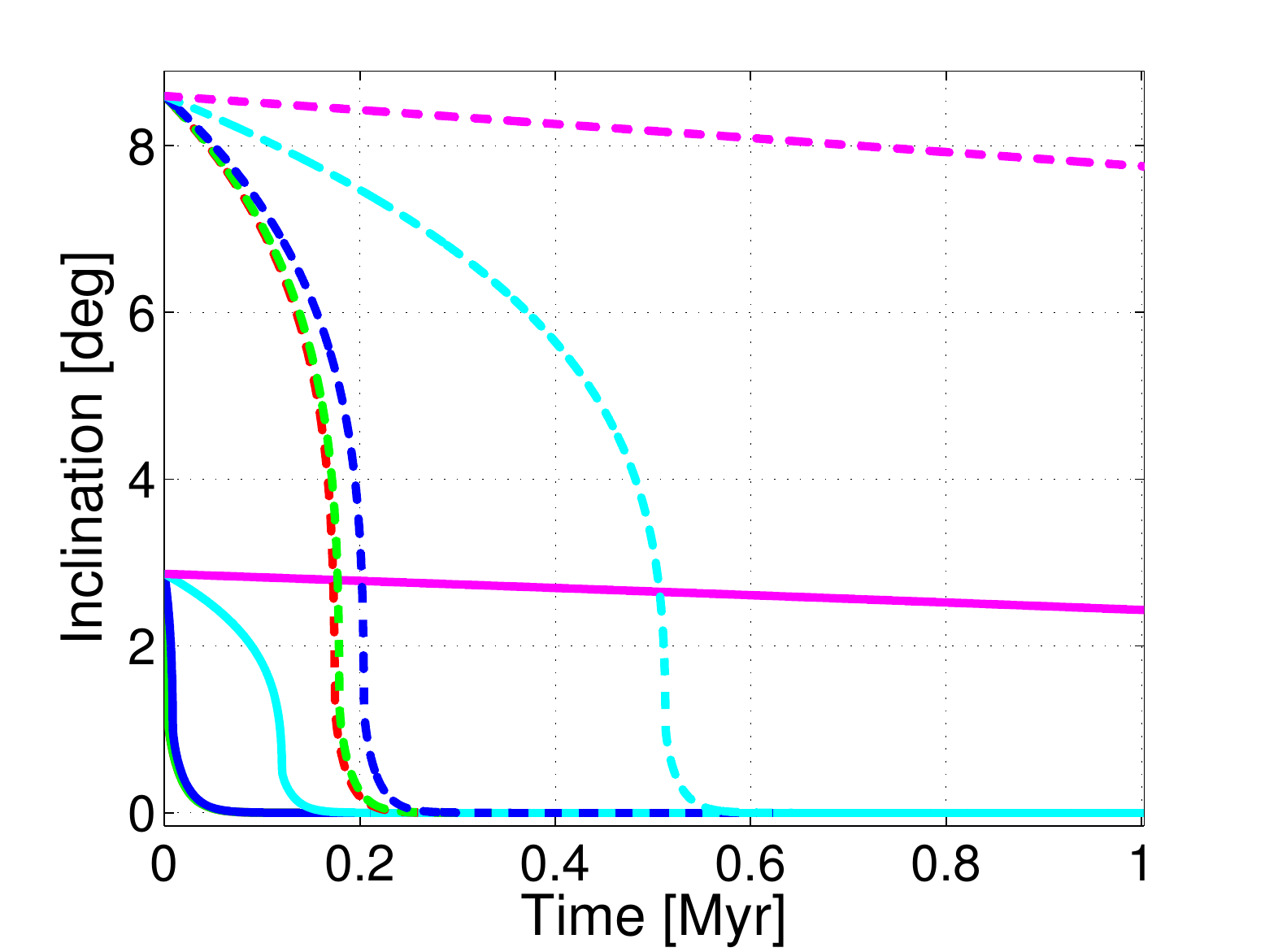}\includegraphics[height=5cm]{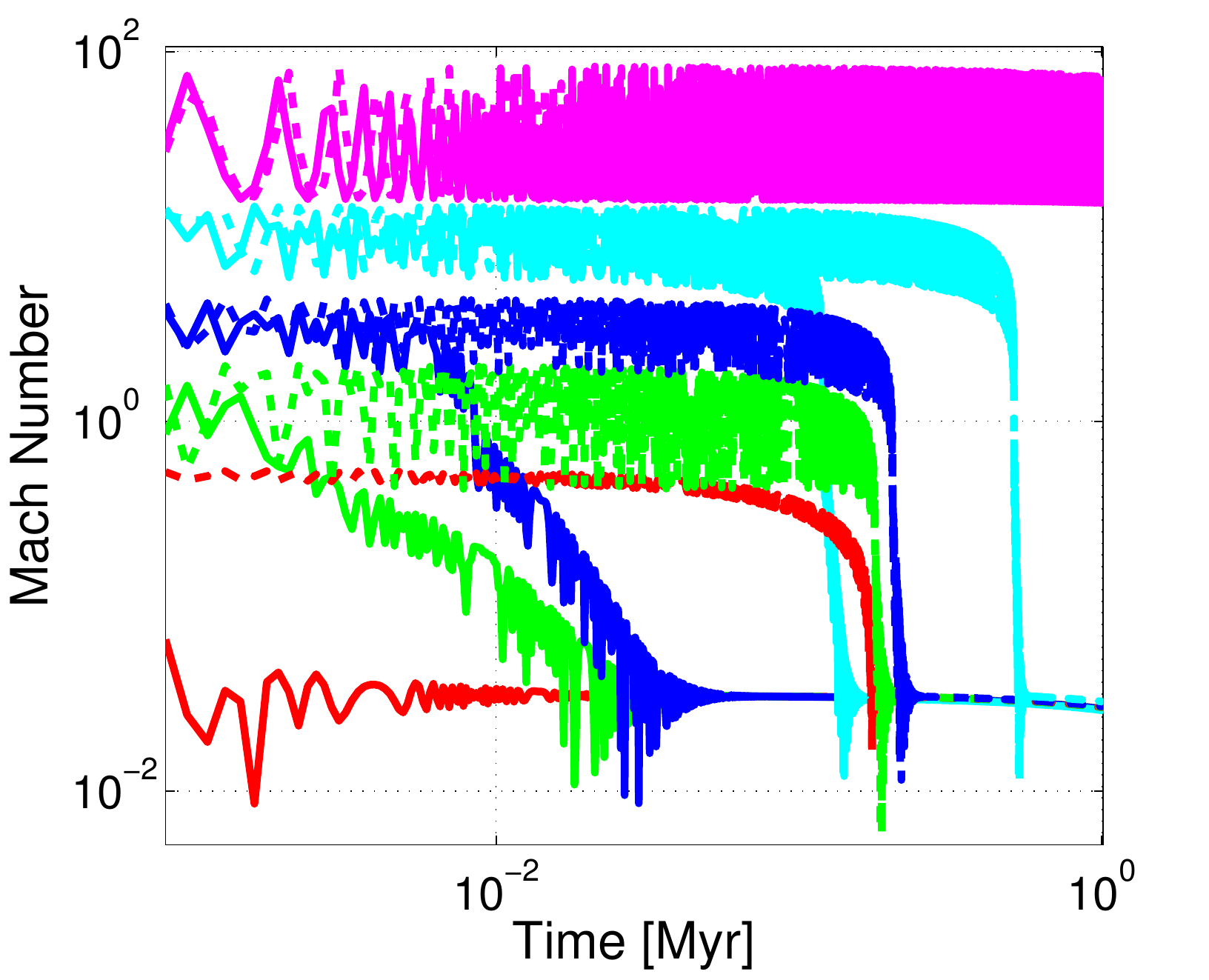} 
\par\end{centering}

\protect\protect\caption{\label{fig:inc1}The evolution of a $2\cdot10^{25}g$ planetesimal
over $1$Myr staring on inclined orbits. Solid lines correspond to
orbits with initially low inclination $I=0.05$ rad, and dashed lines
represent orbits with initially high inclination $I=0.15$ rad. Top
left: Evolution of the semi-major axis. Top right: Evolution of the
eccentricity. Note logarithmic scale. Bottom left: Evolution of the
inclination. Bottom right: Evolution of the Mach number. Note logarithmic
scale.}
\end{figure*}

In figure \ref{fig:inc1} we show the results for initially inclined
orbits. In the first three panels, orbits with low inclination show
a fast decay. 

The average ambient gas density experienced by planetesimals
at high inclination is low compared with the low inclination orbits,
hence GDF becomes less effective. Highly eccentric orbits, $e>0.3$,
share the same fate. In special cases $(e=0.3,\ I=0.15$ rad, dashed
cyan line) the orbit might gain angular momentum and expand, due to
tailwind near the apastron that overcomes the headwind in periastron.

\begin{figure*}
\begin{centering}
\includegraphics[height=5cm]{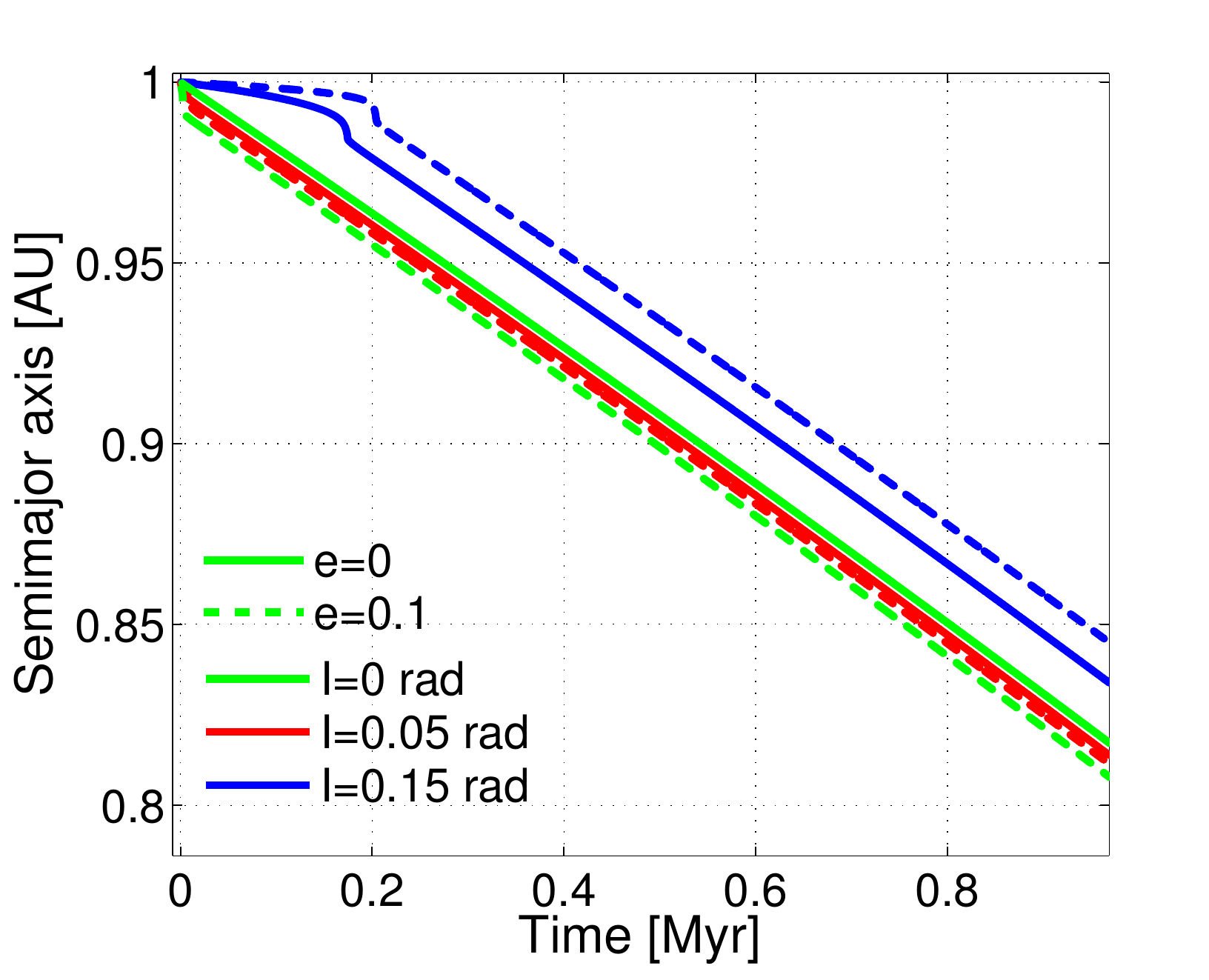}\includegraphics[height=5cm]{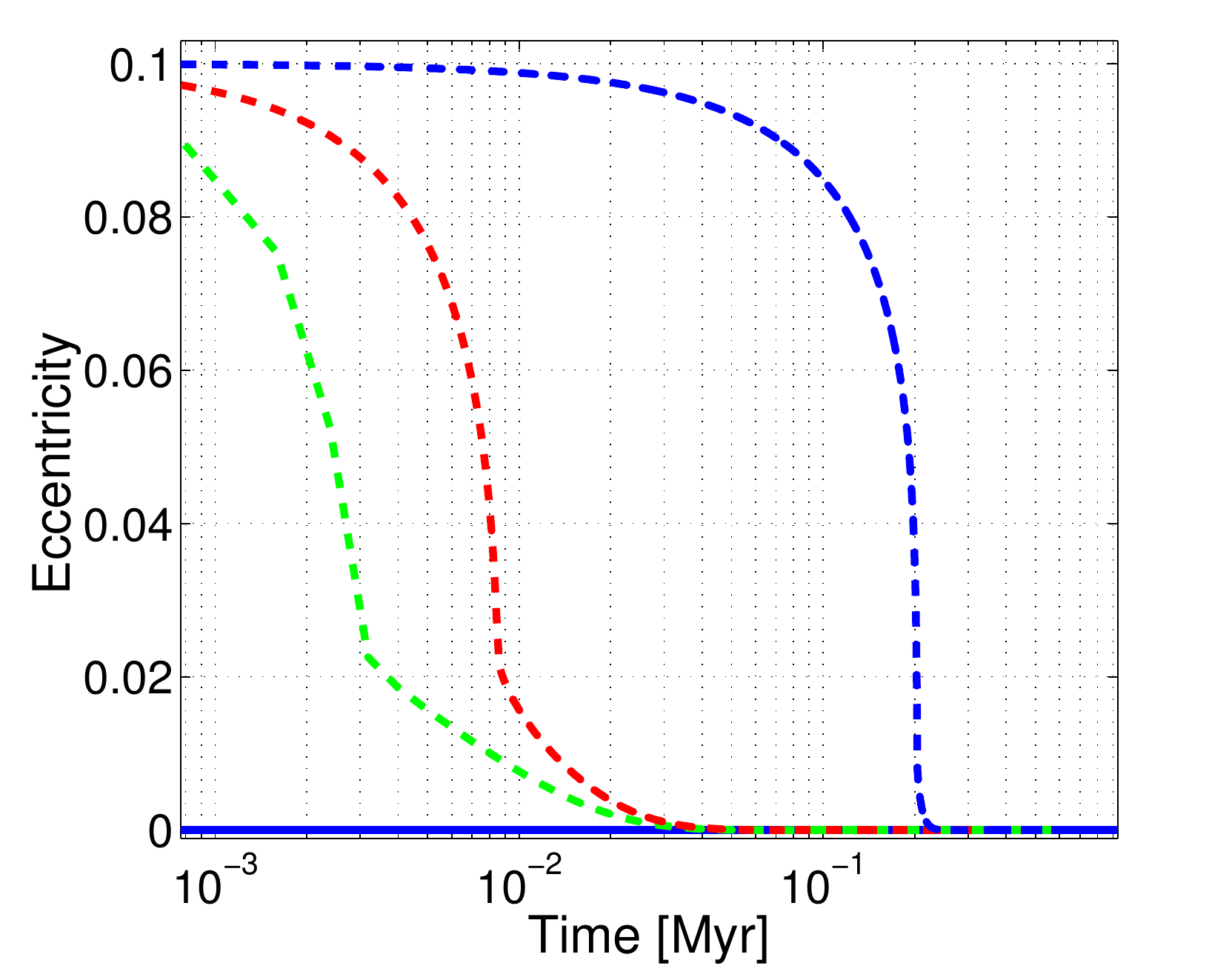} 
\par\end{centering}

\begin{centering}
\includegraphics[height=5cm]{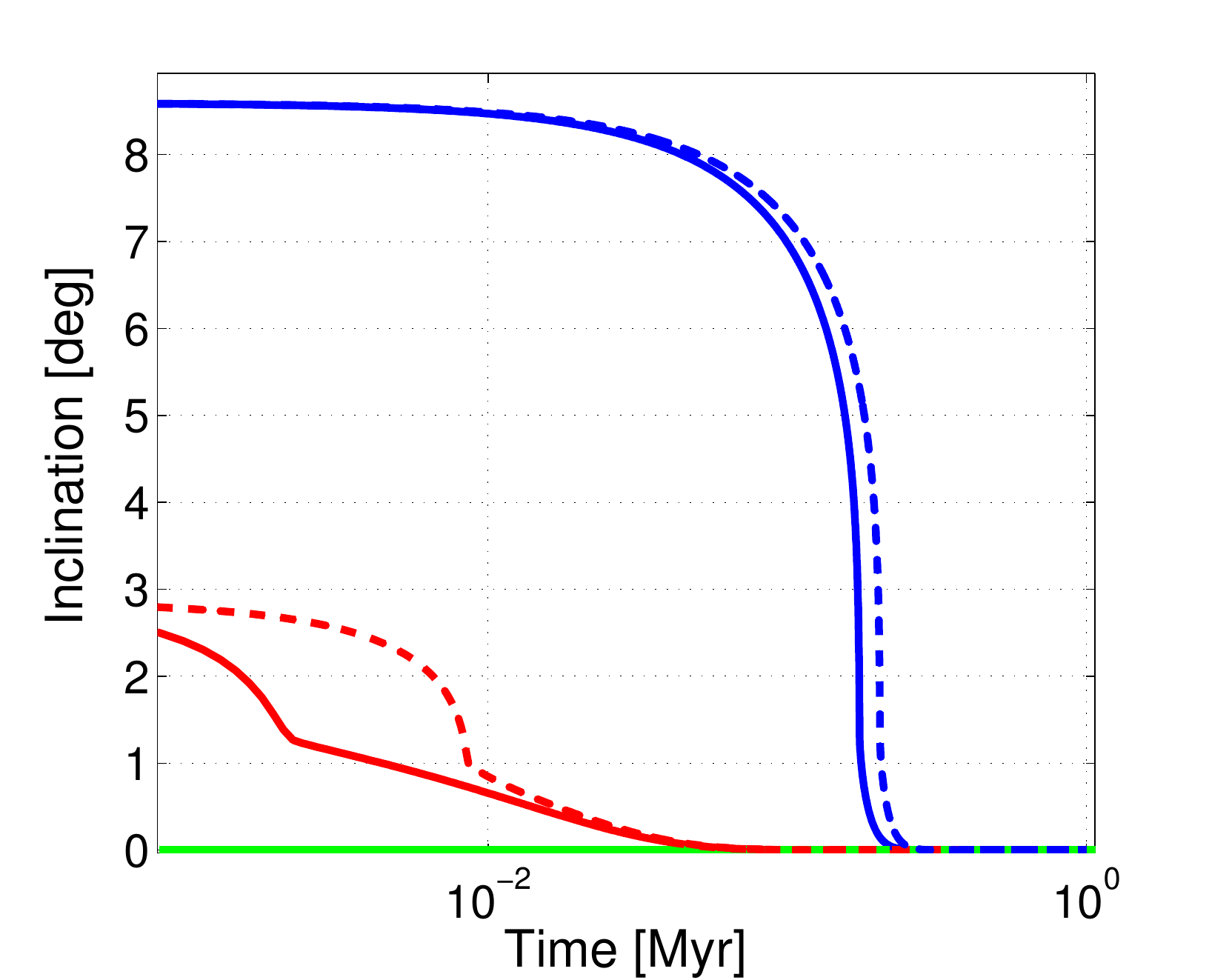}\includegraphics[height=5cm]{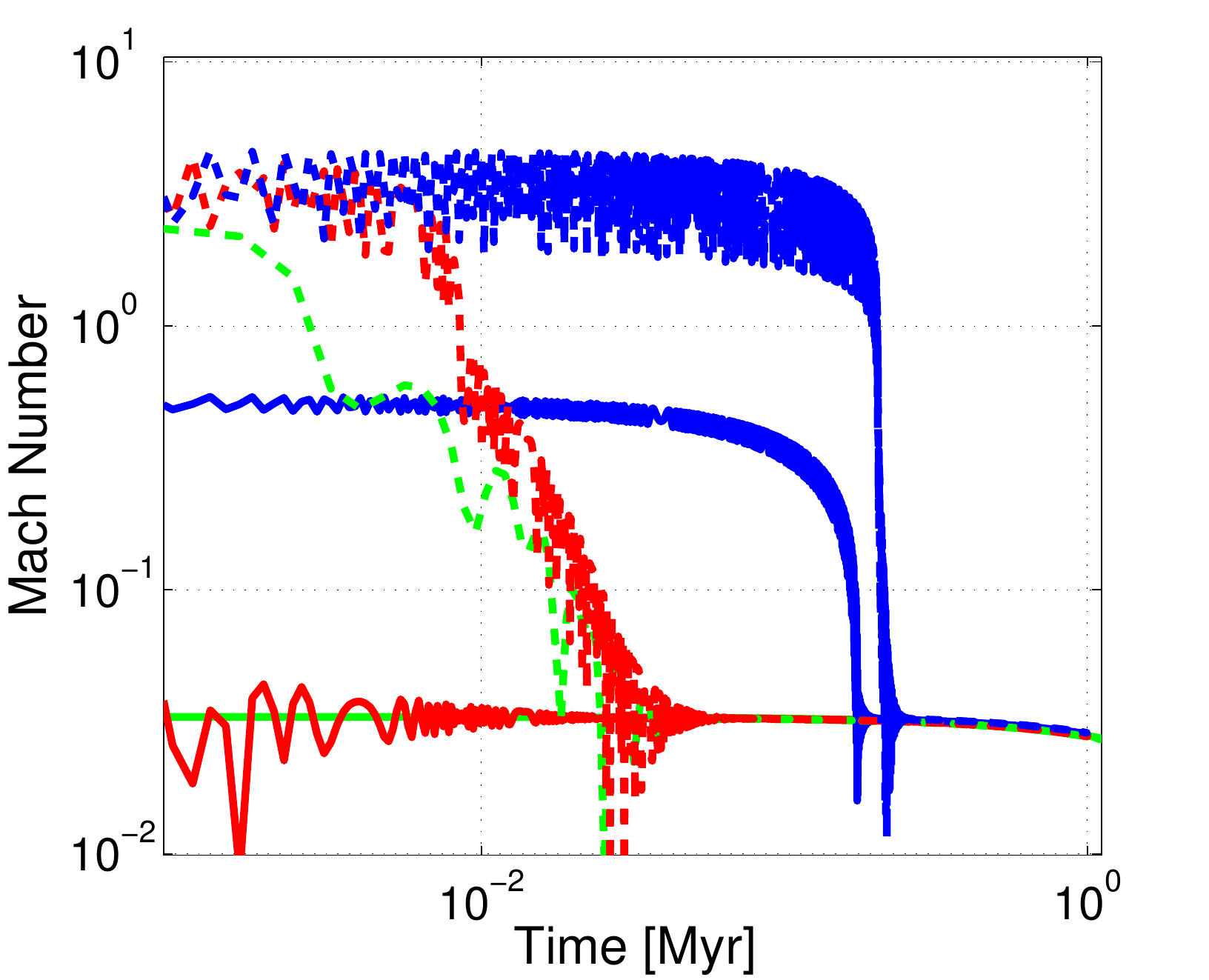} 
\par\end{centering}

\protect\protect\caption{\label{fig:inc2}Comparison of the orbital evolution of a $2\cdot10^{25}g$
planetesimal for a range of initial conditions. Solid lines correspond
to circular orbits, $e=0$. Dashed lines correspond to initially eccentric
orbits $e=0.1$. The color index is green - $I=0$, red - low inclination
$I\sim2H_{0}$, blue - high inclination $I\sim6H_{0}$. Top left:
Evolution of the semi-major axis. Top right: Evolution of the eccentricity.
Note logarithmic scale. Bottom left: Evolution of the inclination.
Note logarithmic scale. Bottom right: Evolution of the Mach number.
Note logarithmic scale.}
\end{figure*}

For low inclination and circular orbit, the inclination drops sharply;
a planetesimal on a circular orbit loses half its initial inclination
already after a few Kyr, comparable to the estimated time-scale in
Eq. \ref{eq:tinc}. Orbits with high inclinations decay much slower,
as expected.

In Fig. \ref{fig:inc2} we compare the evolution of inclined orbits
with those obtained for zero inclination orbits. We consider inclinations
in the range of $I=0-0.15$ rad, comparable with the disk aspect ratio
$\sim2H_{0},6H_{0}$, corresponding to gas density of $1\sigma$ and
$3\sigma$ an the apastron/periastron respectively. The eccentricity
is taken to be either $0$ or $0.1$. As can be seen in the first
panel, the final decay time-scale $\tau_{a}$ is the same in all cases
regardless of inclination and eccentricity, although for high inclination
the decay is somewhat slower at first. On the second panel, we see
that $\tau_{e}$ is different in each case. There is an order of magnitude
difference between orbits with initial inclinations of $I=0$ and
$I=0.05$ rad, and another order of magnitude difference compared
with the initial $I=0.15$ rad case. Thus, $\tau_{e}$ is sensitive
to inclination. In the third panel both the circular and eccentric
orbit decay at the same phase for all inclinations, but the circular
orbits always have lower inclination since the orbit is supersonic
on the eccentric orbits and GDF is less efficient. Only after the
eccentricity becomes negligible does $\tau_{i}$ retains its higher
phase.

\subsubsection{Highly inclined orbits}

In the last section we considered low inclination orbits; however,
observations of the Solar System and other dynamical systems show
an ample of evidence for irregular orbits, with large and retrograde
inclinations. In this sections we explore effects on GDF planetesimals
in such high inclination orbits.

\begin{figure*}
\begin{centering}
\includegraphics[height=5cm]{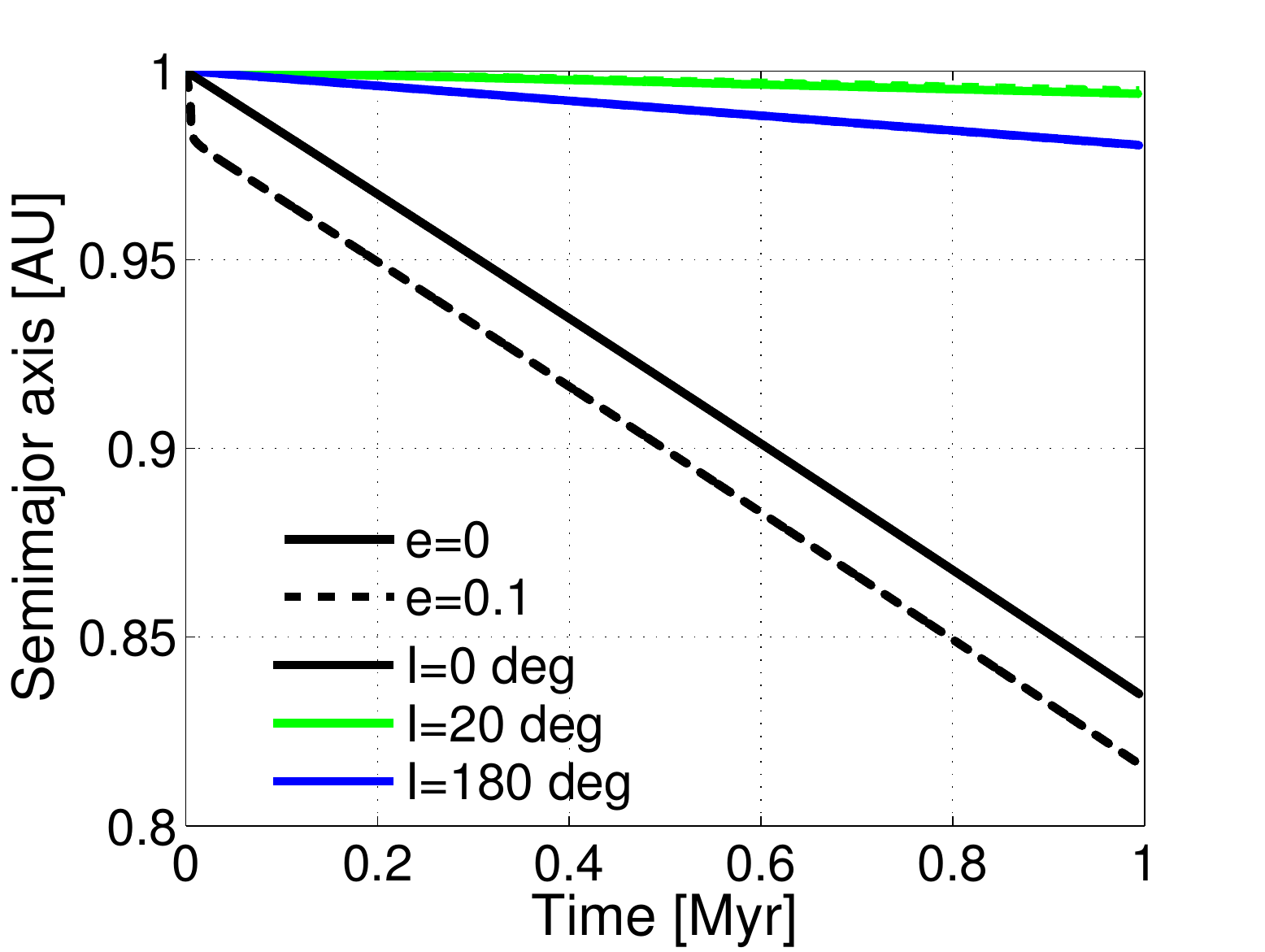}\includegraphics[height=5cm]{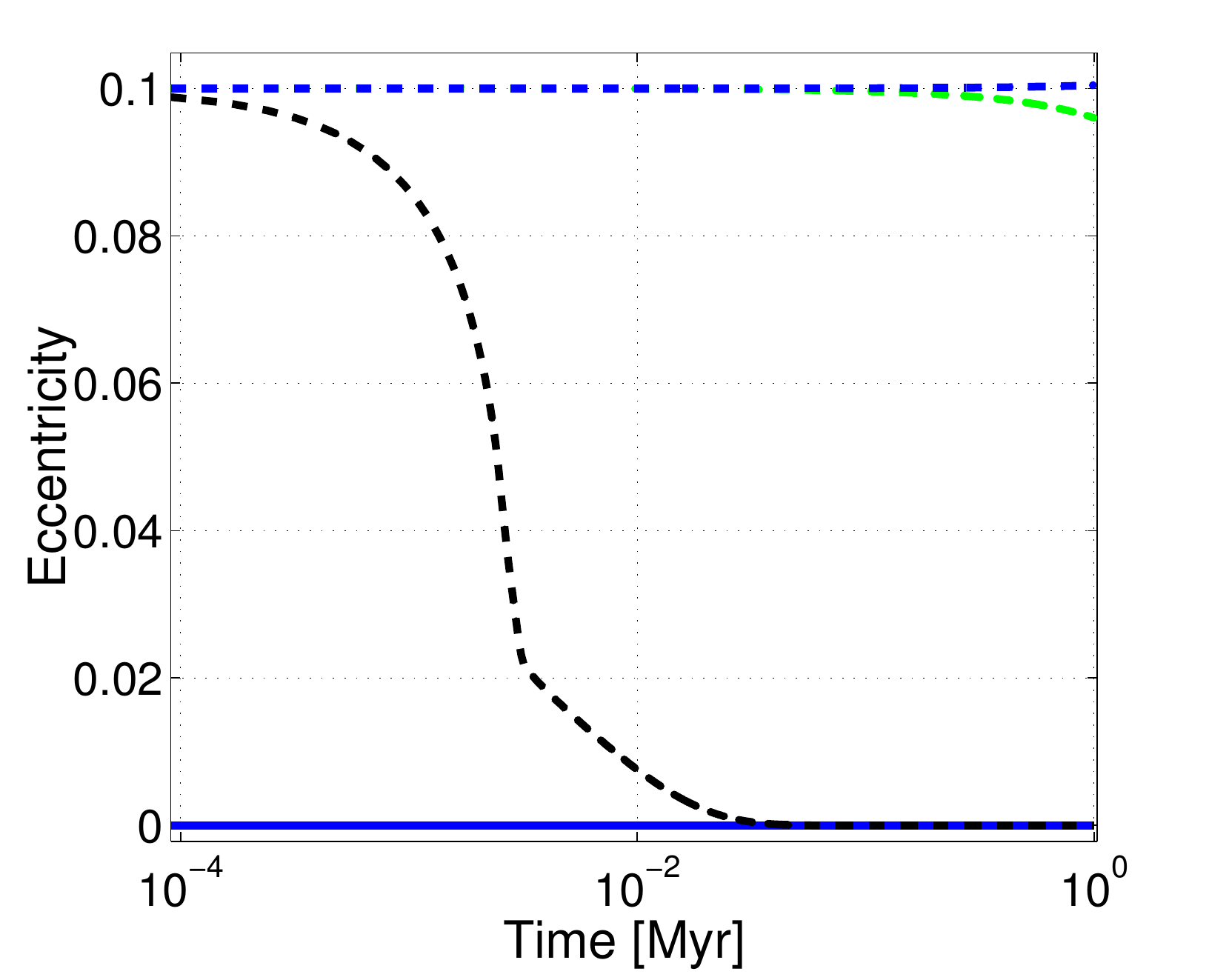} 
\par\end{centering}

\begin{centering}
\includegraphics[height=5cm]{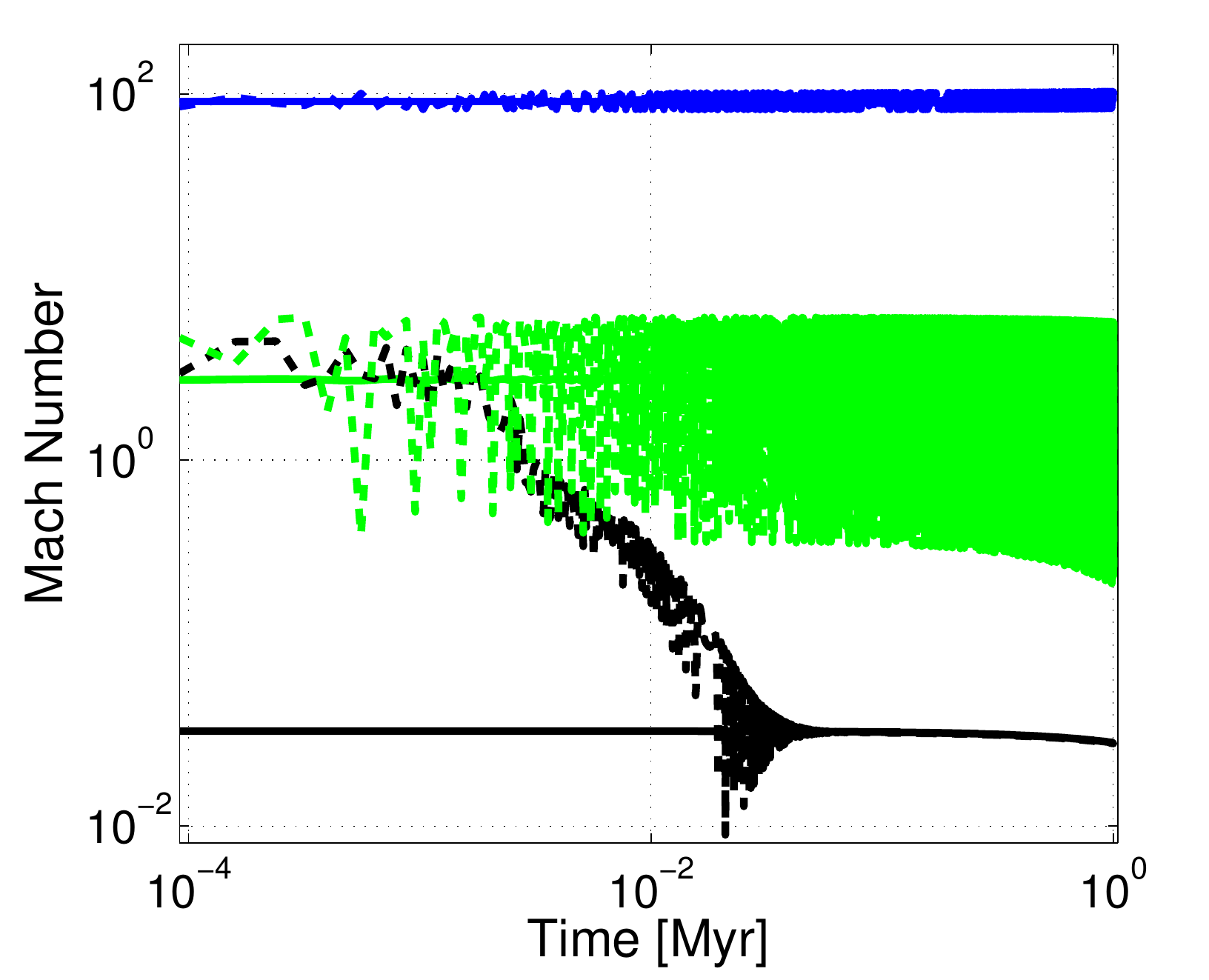}\includegraphics[height=5cm]{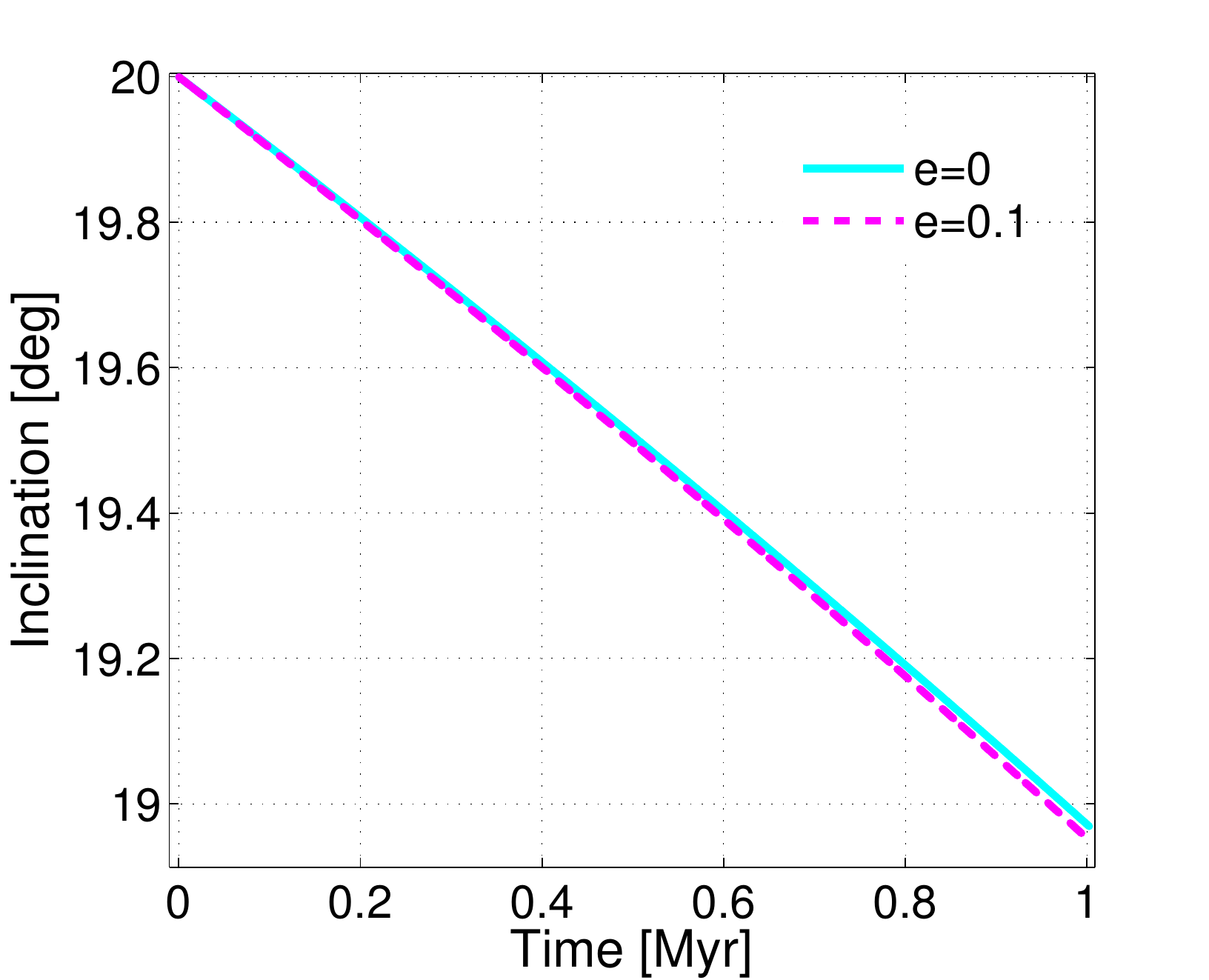} 
\par\end{centering}

\protect\protect\caption{\label{fig:Comparison-between-pregrade-1}Comparison between the evolution
of planetesimals of $m=2\cdot10^{25}g$ with various initial inclinations
and eccentricities. Solid lines indicate circular orbits, dashed lines
indicate eccentric orbits with initial eccentricity of $e=0.1$. Black,
blue and green lines indicate initial inclination of $0,20$ and $180$
deg respectively. Top left: Evolution of the semi-major axis. Top
right: evolution of orbital eccentricity. Note logarithmic scale.
Bottom left: Evolution of Mach number. Note logarithmic scale. Bottom
right: Evolution of inclination. We plot only orbits with initial
inclinations of $20$ deg. The solid cyan line corresponds to an initially
circular orbit, and the dashed magenta line corresponds to an eccentric
orbit with initial eccentricity of $e=0.1$. }
\end{figure*}

Consider for example circular prograde ($I=0$) and retrograde ($I=180^{\circ}$)
orbits. In each case, the dimensionless force is $\mathcal{I}(\mathcal{M})/\mathcal{M}^{2}$.
For prograde orbit, $\mathcal{M}_{prograde}\approx0.033$ and $F_{prograde}=\mathcal{M}_{prograde}/3=0.011$,
while for retrograde orbit, $\mathcal{M}_{retrograde}\approx90$ and
$F_{retrograde}\approx10/90^{2}\approx0.0012$, so $F_{prograde}/F_{retrograde}\approx9$.
We expect that the prograde orbit will decay nine times faster than retrograde
orbit. In Fig. \ref{fig:Comparison-between-pregrade-1} we plot the
evolution of the orbital elements of various initially inclined orbits.
In the top left panel we see that after $1$ Myr, a planetesimal on
a circular prograde orbit ($I=0$ deg, solid black line) has migrated
by $\Delta a_{prograde}=0.17$AU, while the one on a retrograde orbit
($I=180$ deg, blue solid line) has migrated by $\Delta a_{retrograde}=0.02$AU,
hence $\Delta a_{prograde}/\Delta a_{retrograde}\approx0.17/0.02\approx8.5$
, consistent with our expectations.

In the top right panel, only the planetesimal on a co-planar circular
orbit (dashed black line) was able to circularize within $1$Myr of
evolution. In the bottom left panel we only plot orbits wi th initial
inclinations of $20$ deg, since both prograde and retrograde orbits
are co-planar and do not change their inclination. We see that the
changes in inclinations are mild. On the other hand, inclined orbits
with inclination of $\sim9$ deg lost their initial inclinations in
less than a Myr. We conclude that orbits with inclinations higher
than $I\gtrsim4H_{0}\sim12$ deg are not significantly affected by
GDF.

\subsubsection{Disk density profile }

\label{sub:Disk-Density-Profile}

\begin{figure*}
\begin{centering}
\includegraphics[height=5cm]{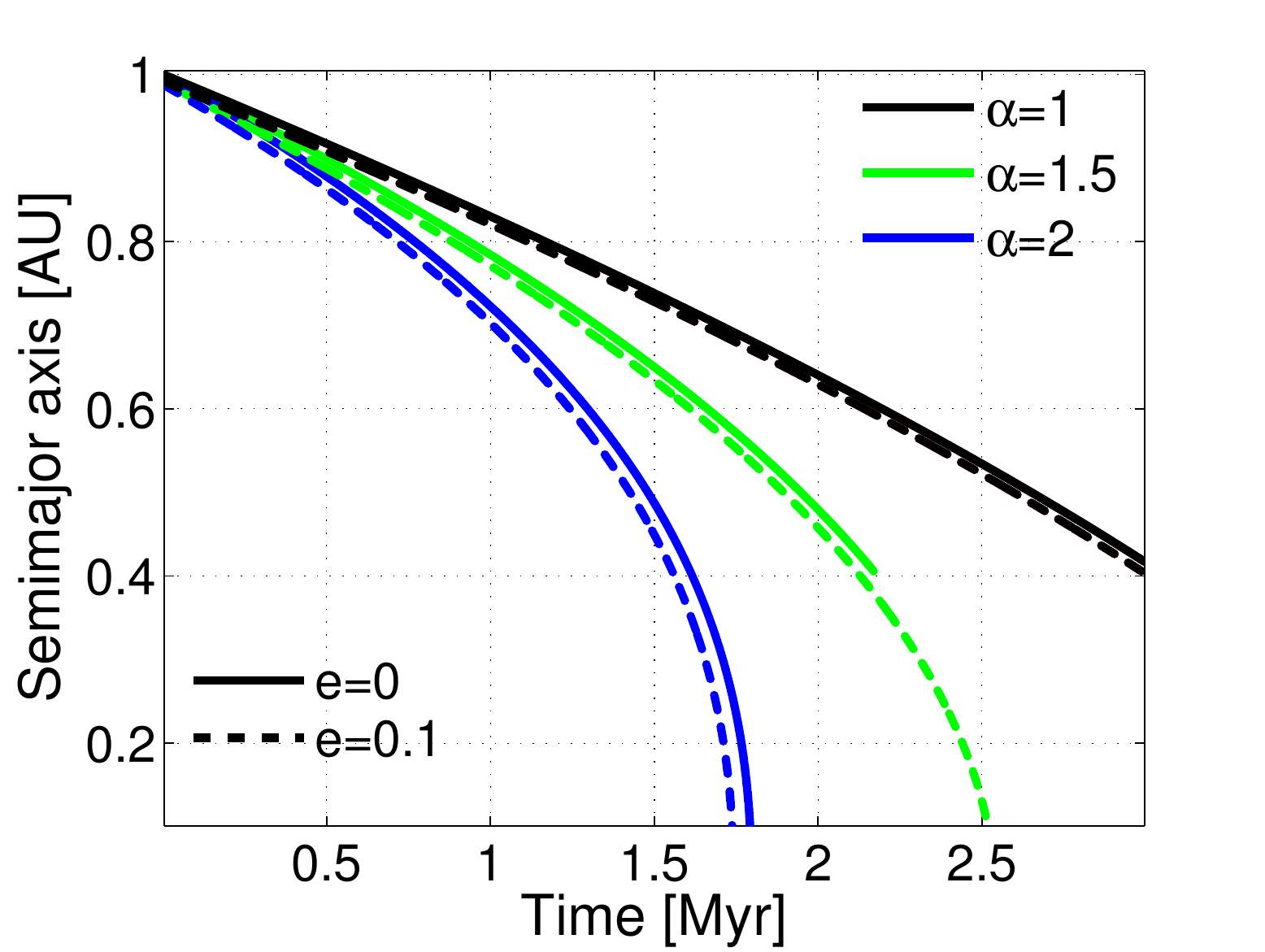}\includegraphics[height=5cm]{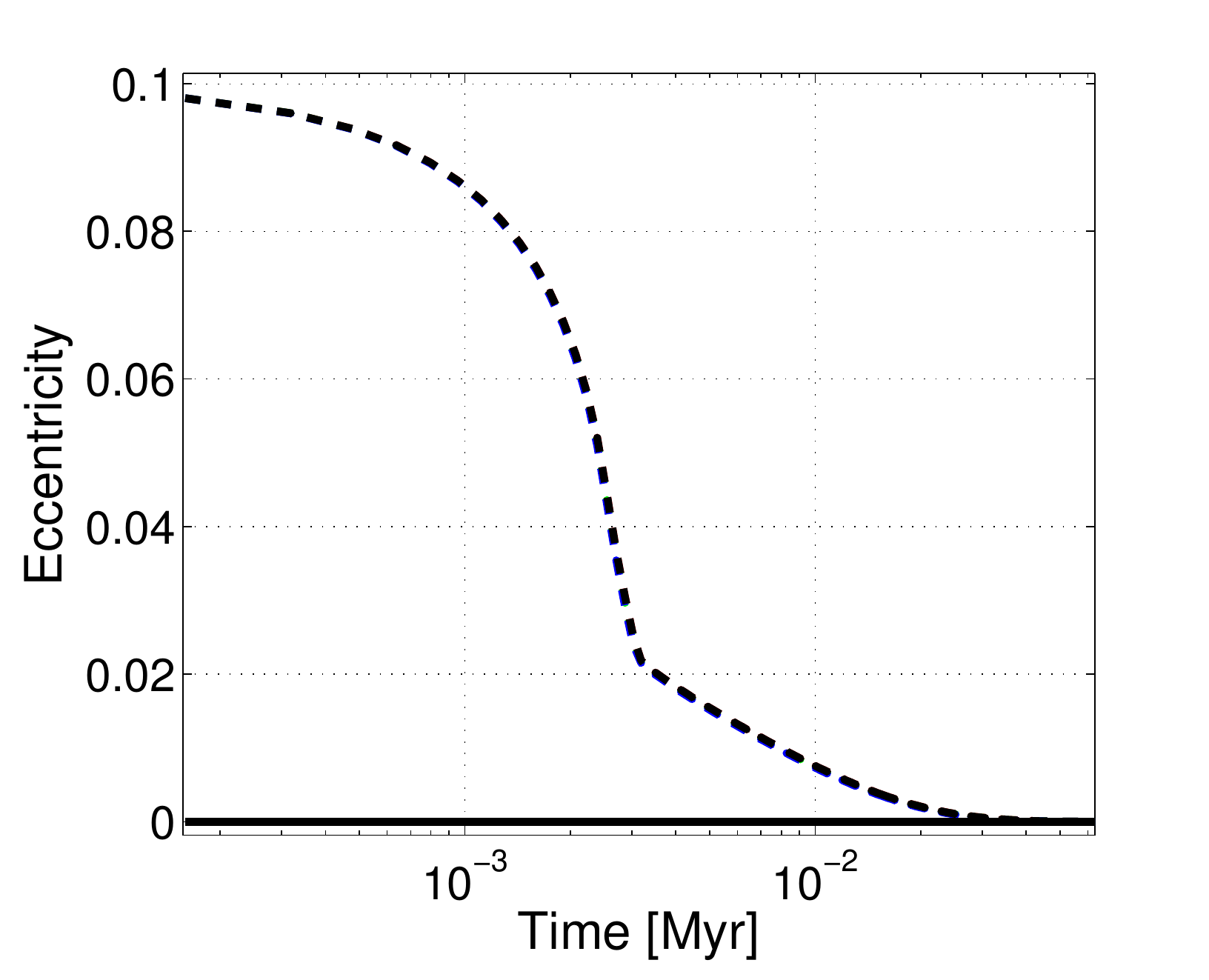} 
\par\end{centering}

\begin{centering}
\includegraphics[height=5cm]{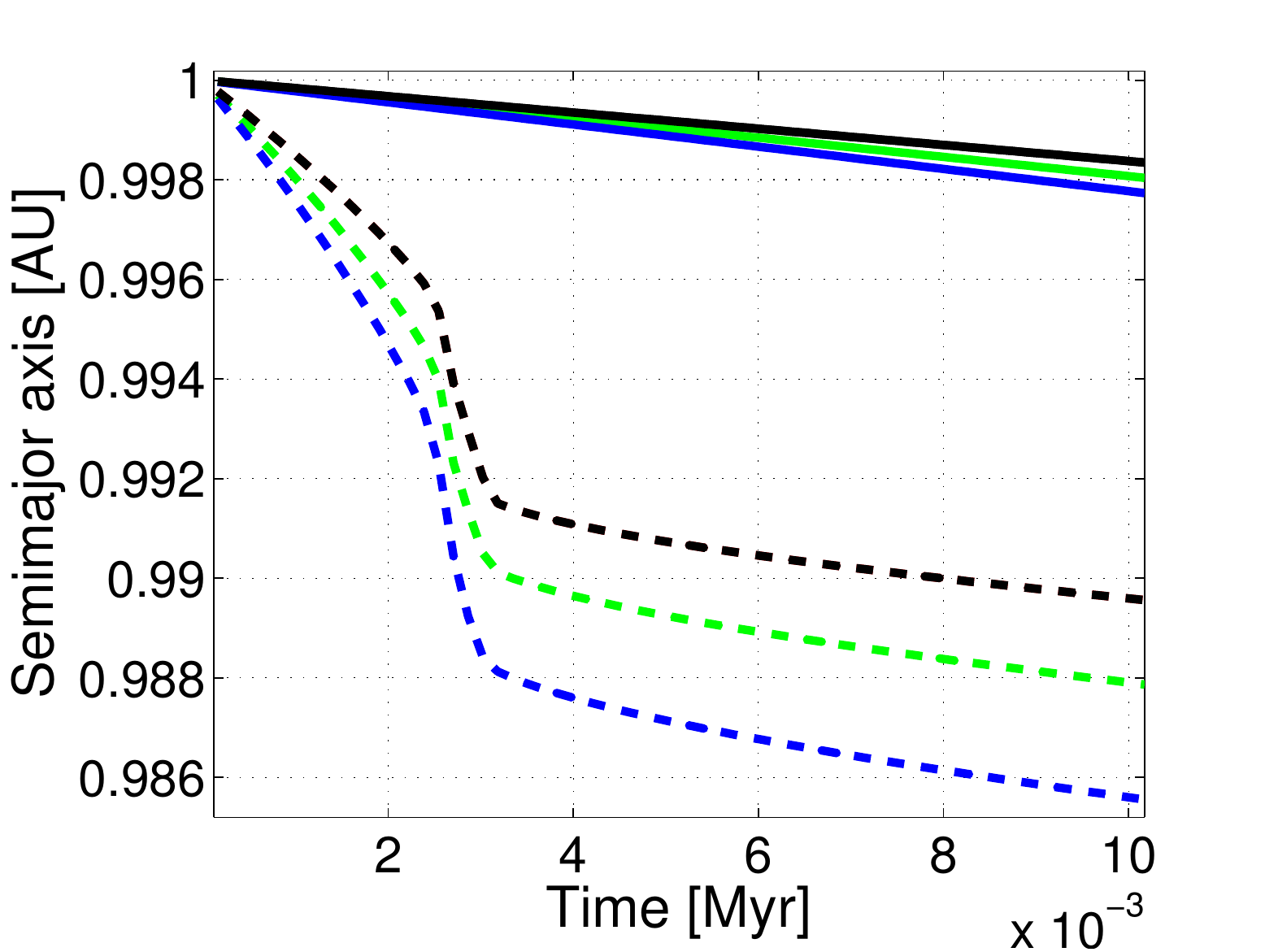}\includegraphics[height=5cm]{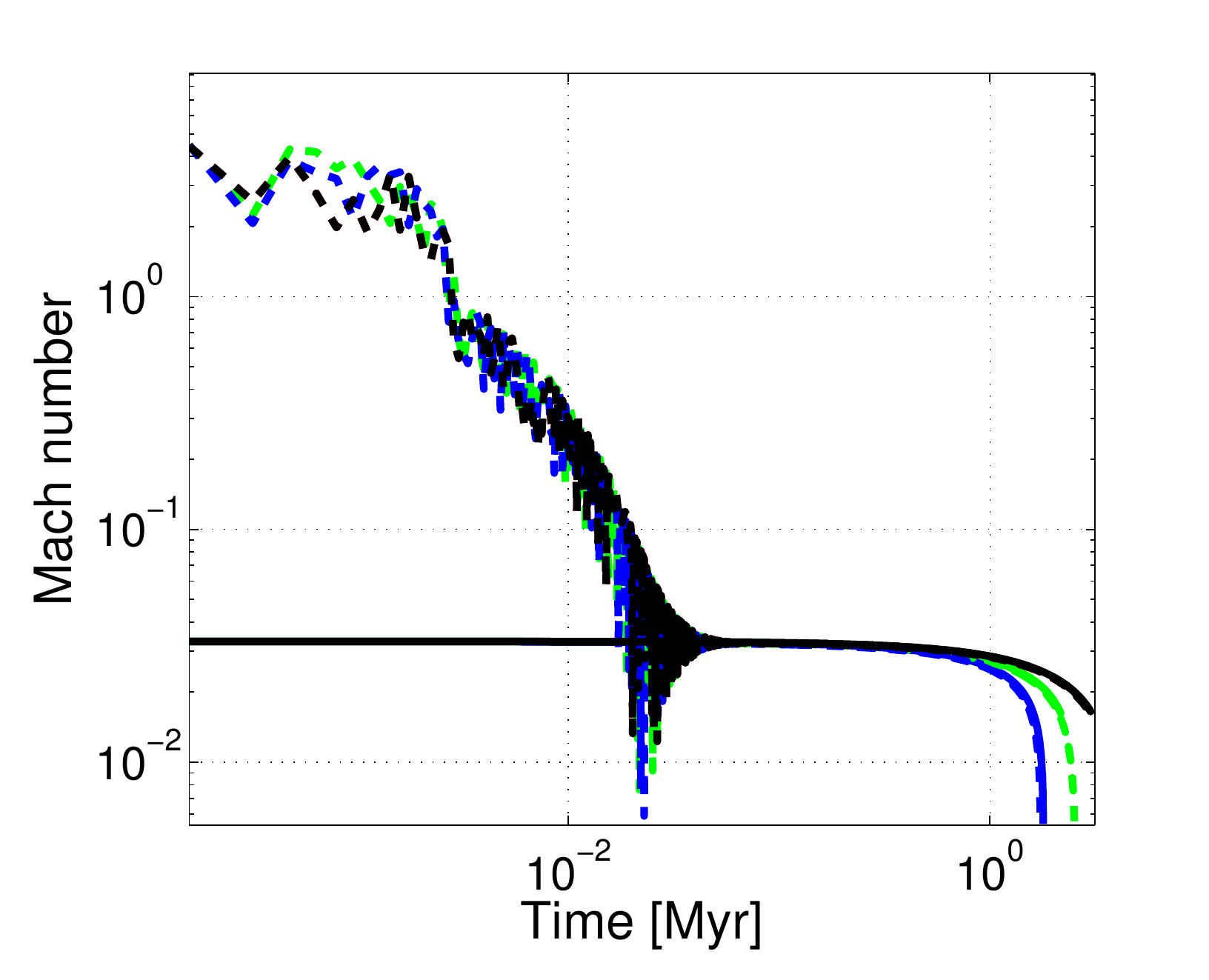} 
\par\end{centering}

\protect\protect\caption{\label{fig:Comparison-between-disk-laws}Comparison between the evolution
of planetesimals of $m=2\cdot10^{25}g$ in protoplanetary disks with
with different disk density profiles normalized at $1$AU. Solid lines
indicate circular orbits, dashed lines indicate eccentric orbit with
initial eccentricity of $e=0.1$. Black, blue and green lines indicate
density profile $\Sigma_{g}\sim a^{-\alpha}$ where $\alpha$ is $1$,
$1.5$ and $2$ respectively. Top left: Evolution of the semi-major
axis. Top right: evolution of orbital eccentricity. Note logarithmic
scale. Bottom left: Evolution of semi-major axis zoomed in on first
$10^{4}$years. Bottom right: Evolution of Mach number. Note logarithmic
scale.}
\end{figure*}

In the previous sections we considered disks with a disk density profile
normalized with $\rho_{g}\sim a^{-9/7-\alpha}$ corresponding to a
surface density profile of $\Sigma_{g}\sim a^{-\alpha}$ with $\alpha=1$.
In the following we consider other disk profiles. In Fig. \ref{fig:Comparison-between-disk-laws}
we run simulations of co-planar orbits with eccentricity of either
$0$ (solid line) or $0.1$ (dashed line). The power law density profiles
considered are for $\alpha=1,1.5,2$, corresponding to the black,
green and blue lines respectively. We have normalized the
density profiles at $1$AU.

As can be seen in the top left panel, migration occurs more rapidly
for larger $\alpha$; the time-scale for migration is shortened by
factor of $\sim3/2$ for $\alpha=1.5$ and $\sim2$ for $\alpha=2$.
The migration time-scale is insensitive to initial eccentricity,
since circularization time-scale is rapid. The fast circularization
is confirmed in the top right panel, where the eccentricity decays
after few Kyrs. Bottom left panel shows the changes in semi-major
axis for the first $10$Kyr. We see that after the initial fast decay
of eccentric orbits, they start to decay in the same manner once they
have been circularized. Only at later times there is difference due
to different ambient density closer to the star. In the bottom right
panel, the changes in Mach number are almost identical, consistent
with the eccentricity decay in the top right panel. In the case of
fixed normalization at $1$AU, $\tau_{a}$ decreases as the density
profile is steeper, with virtually no effect to $\tau_{e}$ .

\begin{figure*}
\begin{centering}
\includegraphics[height=5cm]{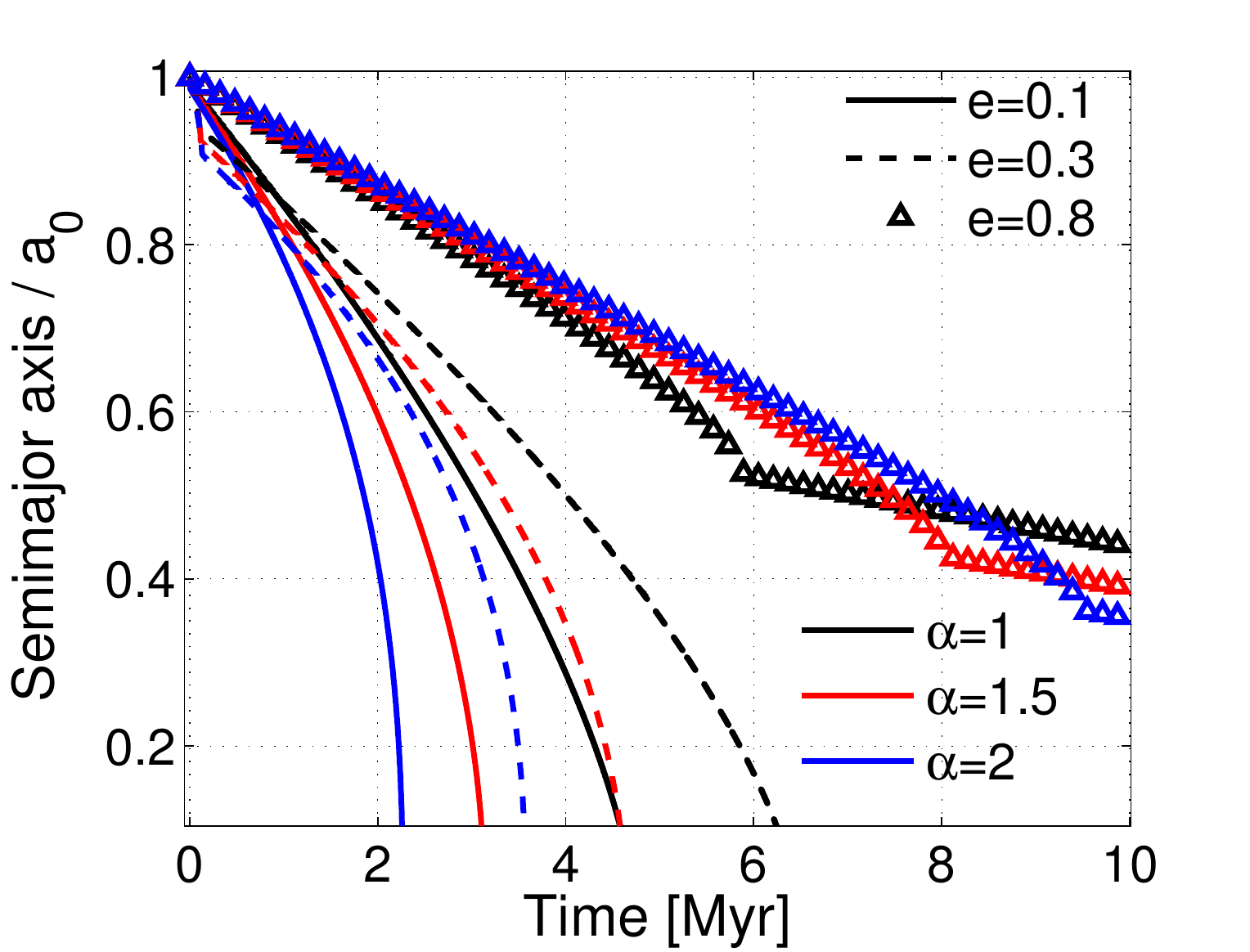}\includegraphics[height=5cm]{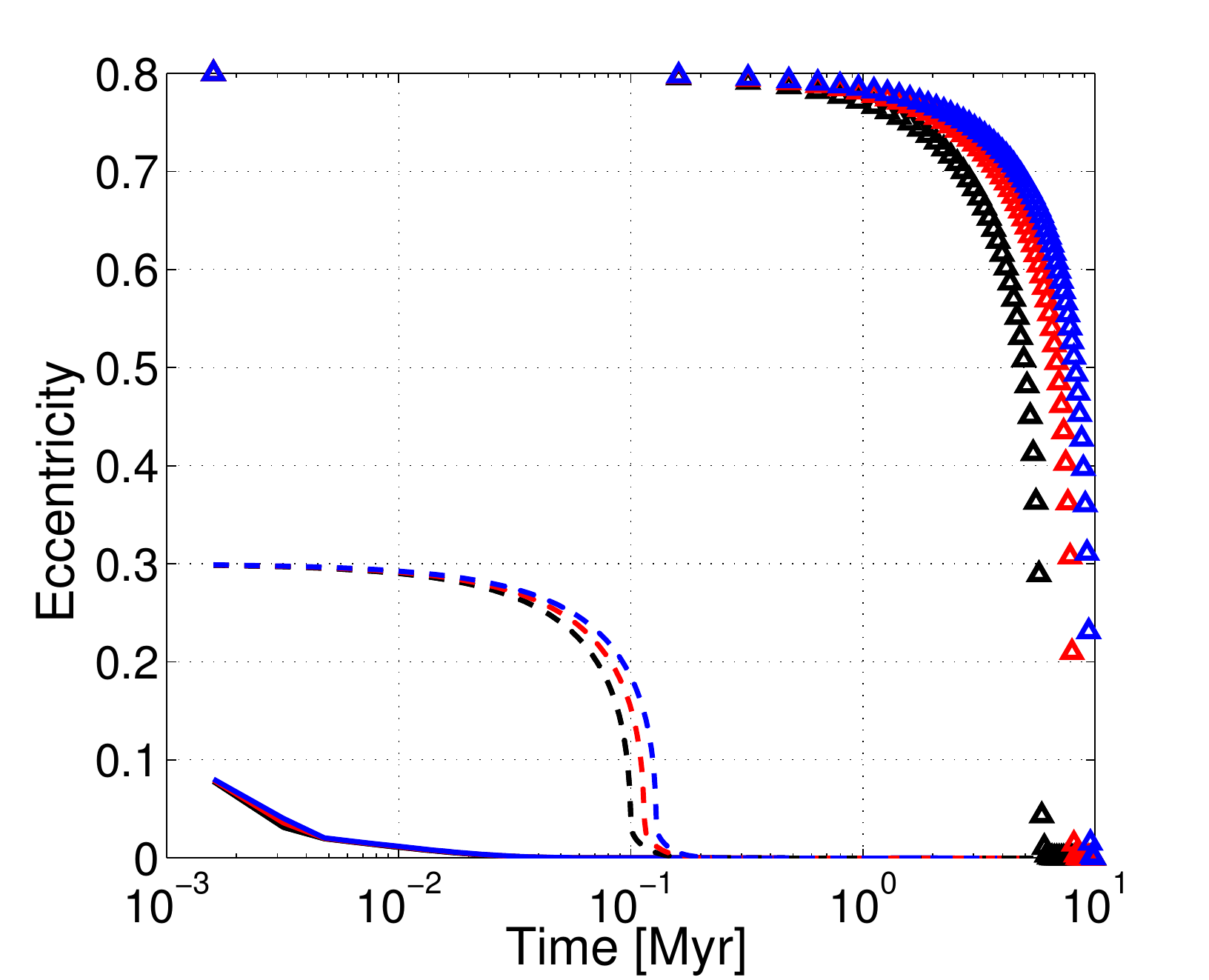} 
\par\end{centering}

\begin{centering}
\includegraphics[height=5cm]{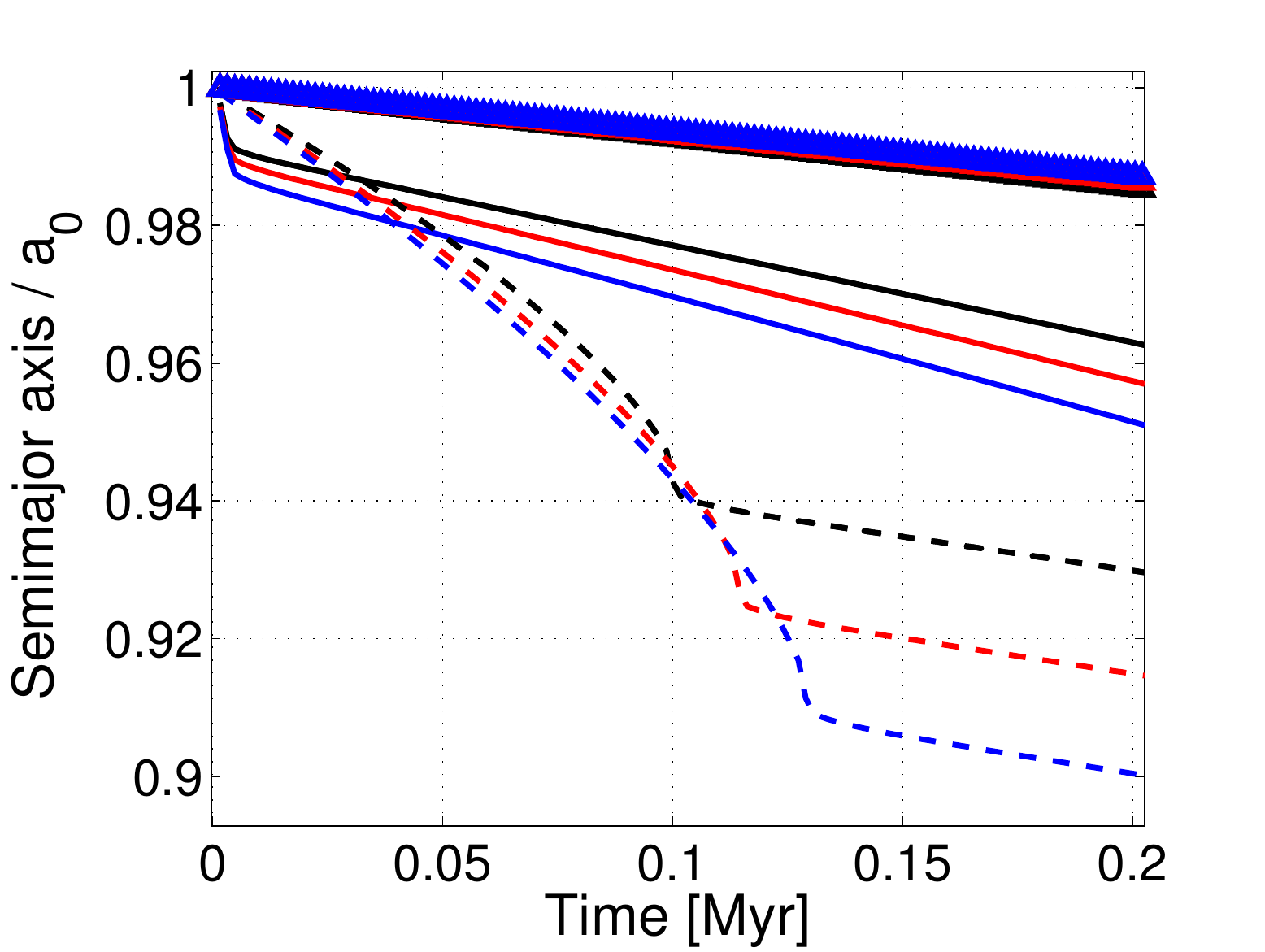}\includegraphics[height=5cm]{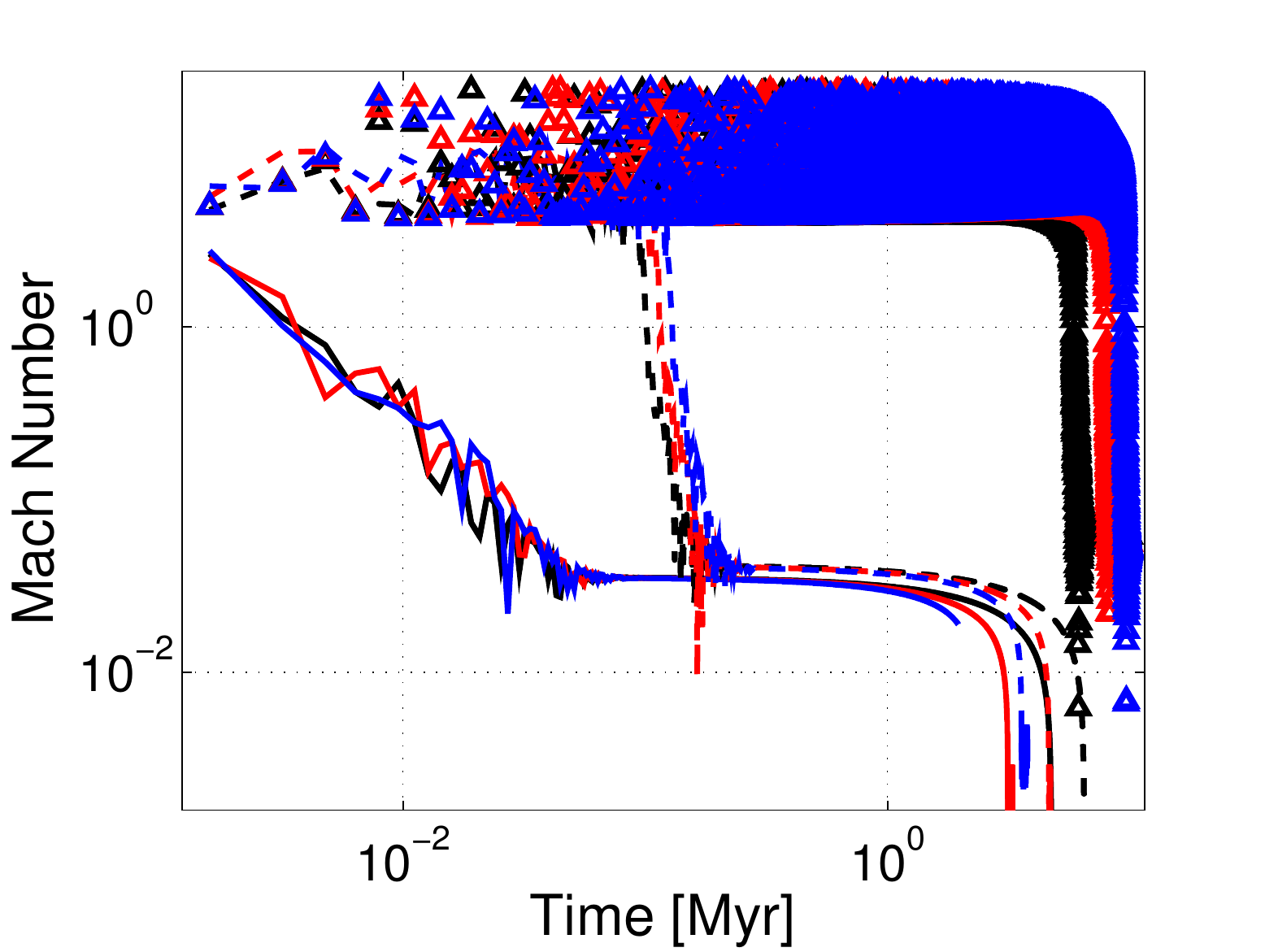} 
\par\end{centering}

\protect\protect\caption{\label{fig:Comparison-between-disk-laws2}Same as Fig. \ref{fig:Comparison-between-disk-laws},
but with normalization at the periastron. Top left: Evolution of the
semi-major axis. Top right: evolution of orbital eccentricity. Note
logarithmic scale. Bottom left: Evolution of semi-major axis zoomed
in on first $10^{4}$years. Bottom right: Evolution of Mach number.
Note logarithmic scale.}
\end{figure*}

In Fig \ref{fig:Comparison-between-disk-laws2}, we run co
planar orbits with $\alpha=1,1.5,2$, initial eccentricities of $e=0.1,0.3,0.8$.
This time, we normalize each density profile to $\rho_{0}$ at the
periastron. On the top left panel we see that the more eccentric orbits
decay slower since their ambient orbit averaged density is lower.
Conversely to the previous case, orbits with different $\alpha$ migrate
and circularize at a different phase.

\subsection{Scaling with planetesimal mass}

\label{sub:pmass}

\begin{figure}
\begin{centering}
\includegraphics[height=6.3cm]{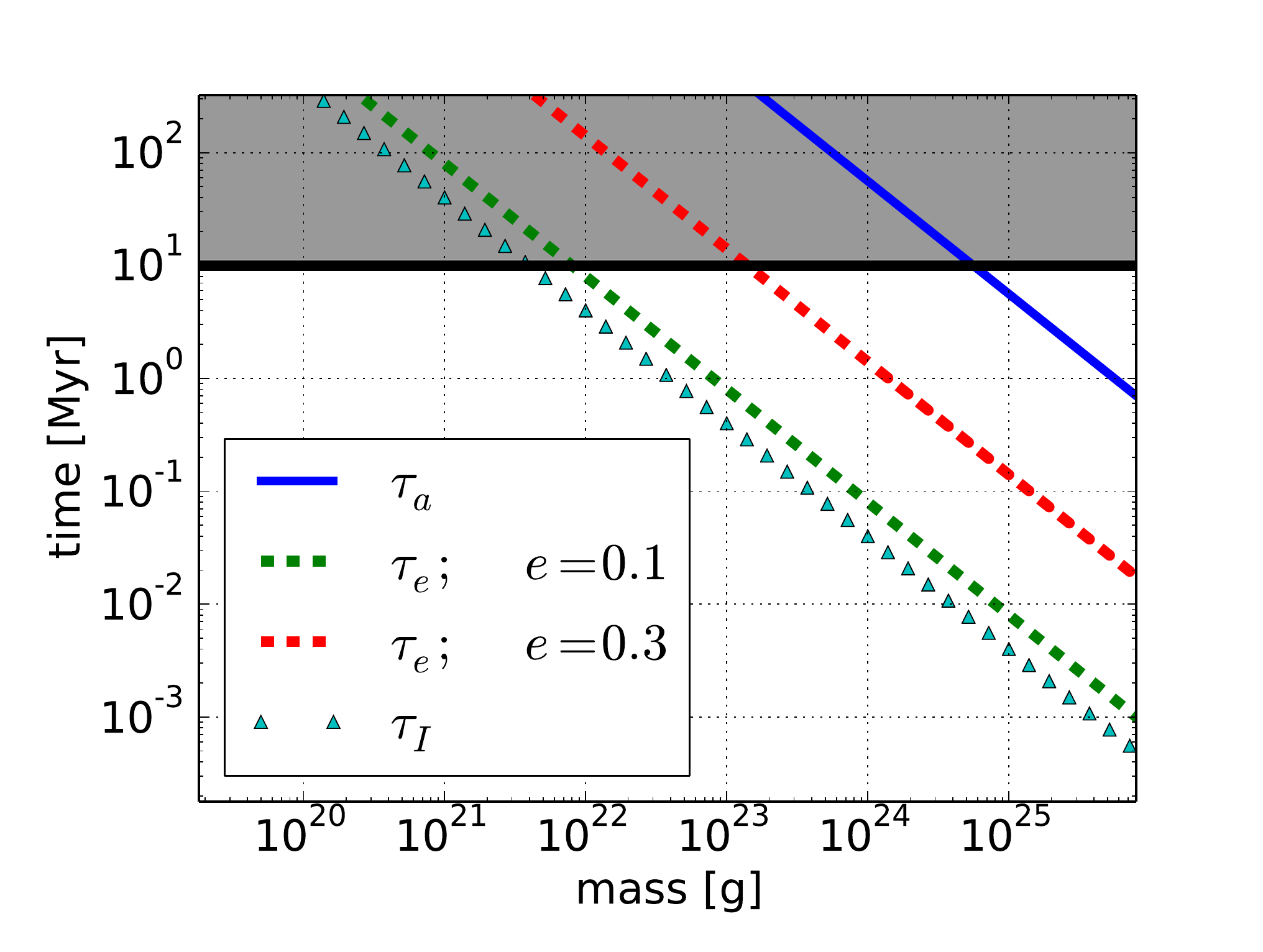} 
\par\end{centering}

\protect\protect\caption{\label{fig:massscaling} Dependence of the time-scales for orbital
evolution obtained from numerical simulations. The shaded region is
excluded due to gas dispersal after $10Myrs$. Solid line indicate
semi-major axis evolution time-scale, $\tau_{a}$(blue). Dashed lines
indicate eccentricity evolution rime scale, $\tau_{e}$, for $e=0.1$
(red) and $e=0.3$ (green). Triangles indicate the inclination evolution
time-scale, $\tau_{I}$, for a circular orbit with initial $I=0.05$
rad. note logarithmic scales.}
\end{figure}

Fig. \ref{fig:massscaling} shows the dependence of the time-scales
for the evolution of the orbital elements on the $\sim m_{p}^{-1}$
mass scaling. We consider time-scales comparable to the gas-disk lifetime
of up to $10Myr$. The slowest evolution time-scale, $\tau_{a}$,
is in found for semi-major axis; only planetesimals more massive than
a few times $10^{24}g$ are significantly affected by GDF. The eccentricity
damping is more rapid. The lower limit for the mass of planetesimals
which are significantly affected (i.e. the damping time-scale
of at least one of their orbital elements, $\tau_{a}$, $\tau_{e}$
or $\tau_{I}$, is comparable to the disk lifetime) is $m\approx8\cdot10^{21}g$
for $e=0.1$ and $m\approx10^{23}g$ for $e=0.3$. Generally $\tau_{e}$
is an increasing function of $e$, but even for high eccentricity
($e\sim0.8$), we still get $\tau_{e}\lesssim\tau_{a}$. The inclination
time-scale, $\tau_{I}$, is by far the most rapid. For low inclination
$I=0.05$ rad we have the lower limit $m\approx4\cdot10^{21}g$ .
Generally $\tau_{I}$ is an increasing function both of initial inclination
and eccentricity. For $I\sim0.15$ $\tau_{I}$ is $\sim100$ times
slower , and starting with $e=0.1$ adds another order of magnitude
to $\tau_{I}$.

It is clear that for single IMPs, GDF damps both inclination and eccentricity
efficiently for most mass ranges. Planetary embryos of mass $>10^{25}$g
also migrate inward on time-scales of a few Myrs.

\section{Discussion}

\label{sec:Discussion}

Before exploring the implications of GDF for the evolution of planetesimal
disks we discuss the various assumptions of which we made use as well
as potential caveats in the approach taken here. We then also compare
the effects of GDF to formulation of type I planetary migration. Finally
we discuss the potential role of GDF in the evolution of protoplanetary
disks.

\subsection{Validity of assumptions and caveats}

\label{sub:Caveats}

\textbf{Accretion}:

We have shown in section \ref{sub:formulation} that gas accretion
is negligible for planetesimal radius $R\lesssim1000$km. In section
\ref{sec:GPI} we have shown that gas drag is negligible for $R\lesssim500$km.
Thus, for intermediate mass planetesimals both gas drag and accretion
are negligible. This intermediate regime therefore complements \citet{2011MNRAS.416.3177L}'s
analysis, where the dominant force considered was due to accretion.
For larger radii and fully formed planets accretion is the dominant
force and one should use models with accretion as the dominant drag
force (e.g. \citealp{2011MNRAS.416.3177L}, \citealp{2014AA...561A..84L}).

\textbf{Linear regime}:

\citet{2010ApJ...725.1069K} has studied the non-linear regime of
GDF. The non-linearity parameter is 
\[
\mathcal{B}=\frac{Gm_{p}}{c_{s}^{2}R(\mathcal{M}^{2}-1)}=\frac{r_{B}}{2R(\mathcal{M}^{2}-1)}
\]
For low Mach numbers it reduces to the condition for accretion. Hence
we conclude that for planetesimals of radius $R\lesssim1000$km the
linear regime considered here is applicable. For larger embryos non-linear
effects are non-negligible, where non-linear effects tend to decrease
the GDF force by a factor of a few \citep{2010ApJ...725.1069K}.

\textbf{Time dependent 3D geometry vs. steady state 2D geometry}:

\citet{1999ApJ...513..252O}'s original derivation made use of time
dependent perturbation theory, where the perturbation is turned on
at time $t=0$. In a time dependent analysis there is a non-zero force
in the subsonic regime, however the surprising result is that this
force is time independent. The reason is that contributions from large
distances (the far field) are not negligible. However, for disks,
the emitted sound waves eventually reach the vertical edge $h$ of
the disk after time $t\sim h/c_{s}=1/\Omega\sim1$yr. Hence, after
one orbital period the waves cannot propagate further and the contribution
from the far field decreases until it becomes negligible.

In order to tackle the problem, \citet{2011ApJ...737...37M} solved
the equations of motion in a slab geometry. They introduced an averaged
potential at the scale height of the disk, i.e. $\Psi=-Gm_{p}/\sqrt{x^{2}+y^{2}+(\gamma h)^{2}}$
where $\gamma$ is of order unity factor. They decomposed the solution
into Fourier components and searched for steady state solutions. Moreover,
they showed that time dependent solutions decay as $\sim t^{-1}$,
and therefore after $t\sim1$yr the wake persists in a steady state.

Nevertheless, such a steady state might never be reached. \citet{2011ApJ...737...37M}
note that the assumption for a steady state is only marginally satisfied.
Moreover, there is uncertainty in the 2D approximation, and the force
might be altered by a factor of a few. In addition, protoplanetary
disks tend to be turbulent, which may significantly affect the perturbation
evolution. The turbulence is parametrized by the turbulent viscosity
$\nu=\alpha c_{s}h$ where $\alpha$ is the Shakura-Sunyaev parameter
\citep{1973AA....24..337S}, not to be confused with disk density
scaling power law. Consider a Kolmogorov distribution, i.e. the flow
consists of self-similar eddies, and energy cascades from the largest
eddy to the smallest one, where it is dissipated by molecular viscosity
\citep{2010AREPS..38..493C}. In this case, the largest eddies are
of order $l_{0}\sim h$, and the eddy turnover time-scale is $t_{0}=l_{0}/v_{0}=1/\Omega$.
where $v_{0}\sim c_{s}$ is the velocity of the largest eddy. After
a few $t_{0}$, the gas is well mixed and the density wave starts
to propagate again from the planetesimal outwards. This is equivalent,
in some sense, to restarting the problem. Since $t_{0}\lesssim1/\Omega$,
the steady state might not be reached, and the pressure wave propagation
is restarted every few$\times$ $t_{0}$.

For smaller eddies, the characteristic eddy turnover time is\footnote{It is worth noting that the proportionality constant for $l_{0}$
and $v_{0}$ might be different from unity. \citet{2001ApJ...546..496C}
take $l_{0}=\sqrt{\alpha}H_{0}$ and $v_{0}=\sqrt{\alpha}c_{s}$ } $t_{l}\sim(l/l_{0})^{2/3}t_{0}$. Thus, for a wake with characteristic
length $l=c_{s}t$, the relevant turnover time-scale corresponding
to its length is $t_{l}/t=(\Omega t)^{-1/3}$. For $t\ll1/\Omega$,
the perturbation is not affected by the eddy current, and only for
$t\sim1/\Omega$ the turbulent current of the largest eddy destroys
the wake, and one therefore needs to consider a time-dependent evolution
following Ostriker.

Another possible reason for the steady state not being reached
is that the planetesimals come back to (nearly) the original position
after one orbit. For subsonic regime the gas has enough time to rearrange
itself, but for super-sonic regime subsequent perturbations of the
disk are possible. \citet{2007ApJ...665..432K} find the number of
interactions with the wake increases monotonically with larger Mach
numbers.

Another attractive feature in \citet{1999ApJ...513..252O}'s linear
theory, is that it does not deal with viscosity and dissipation, which
would affect the model studied by \citet{2011ApJ...737...37M}, since
these are second order effects. Our approach is therefore complimentary
and consider the time dependent approach following Ostriker.

If $\alpha$ is low, and the disk is laminar, we can compare between
both models. Denoting the GDF forces \citet{1999ApJ...513..252O}'s
3D model and \citet{2011ApJ...737...37M}'s 2D model as 
\begin{equation}
F_{3}(\mathcal{M})=F_{0}\mathcal{I}(\mathcal{M})
\end{equation}
and 
\[
F_{2}(\mathcal{M},\alpha)=F_{0}\left\{ \begin{array}{cc}
\alpha/8 & \mathcal{M}<1\\
F_{0}/2 & \mathcal{M}>1
\end{array}\right.
\]

respectively, where $F_{0}=4\pi G^{2}m_{p}^{2}\rho_{g}/v_{rel}^{2}$
(see Eqs. (30) and (41) in \citealp{2011ApJ...737...37M} for details).
We estimate the ratio between both models 
\[
\frac{F_{3}}{F_{2}}=\left\{ \begin{array}{cc}
\frac{8}{3\alpha} & \mathcal{M}<1\\
2\ln\frac{Vt}{r_{min}}+\ln(1-\mathcal{M}^{-2}) & \mathcal{M}>1
\end{array}\right.
\]

For a subsonic perturber, with $\alpha\sim10^{-2}$, the difference
is 2-3 orders of magnitude. The situation gets better for supersonic
perturber, where the second term is negligible, and the difference
is $\sim2\ln Vt/r_{min}$. The Coulomb logarithm is not well defined,
but here $\ln\Lambda\sim\ln h(a)/r_{min}\sim9-10$, so the difference
becomes smaller, only one order of magnitude. We simulated
planetesimals of $2\cdot10^{25}g$, or with radius of $1000km$. The
minimal radius is the minimum of either the physical radius of the
planetesimal, the Bondi radius or the non-linearity radius. For IMPs
$r_{min}$ is the physical size the planetesimal, of the order of
$1000$km. The two models are therefore marginally compatible in
supersonic regime, where as a much more significant difference is
expected in the subsonic regime.

\textbf{Shear}:

The derivation for linear regime is valid for homogeneous gas. In
reality, Keplerian disks have differential shear. \citet{2011ApJ...737...37M}
suggest that the shear is negligible if the distance from the planet's
semi-major axis to the \emph{instantaneous co-rotation radius, }defined
by $a_{C}=a(1-e)/(1+e)$ is larger the the relevant length scale,
i.e. $|a_{C}-a|\gtrsim h$, or $e\gtrsim2H_{0}$. However, they neglected
the pressure gradients since they were initially interested in supersonic
orbits. In our case, for circular orbit spiral density wave
appears where the relative velocity is supersonic and the effective
Lindblad resonances accumulate \citep{1993ApJ...419..155A}, so the
morphology of the wave will be distorted on scales comparable to the
scale height $\sim H_{0}a=h$. The overall force will therefore
potentially differ by a factor of a few.

\subsection{Connection to type I migration}

\label{typeI}

In section \ref{sec:GPI} we have compared between aerodynamic
gas drag and GDF. As mentioned earlier, these forces originate from
essentially different physical processes. Aerodynamical gas drag originates
from difference in the pressure due to changes in the flow around
the body. It is proportional to the area and depends on geometry.
GDF, on the other hand originates from gravitational interactions
between the massive body and the ambient gas, and proportional to
the mass (squared) of the body. Aerodynamical gas drag acceleration
decreases with increasing mass, while GDF acceleration increases with
increasing mass. The comparison between gas drag and GDF is straightforward.

In the case of GDF and type I migration, however, the comparison
and the connections are less trivial. Both approaches deal with gravitational
interactions between the planetesimal and the gaseous disk.

Up to dimensionless factors\footnote{The dimensionless factors are not necessarily order unity.
Some of them could be very small, e.g. for the subsonic regime the
dimensionless factor is proportional to $\sim\mathcal{M}^{3}$}, the GDF torque is 
\[
T_{gdf}\sim a\frac{G^{2}m_{p}^{2}\rho_{g}}{c_{s}^{2}}
\]
while the type I migration torque (e.g. \citealp{2013apf..book.....A,2002ApJ...565.1257T})
is 
\begin{equation}
T_{migration}\sim\Sigma_{g}\Omega^{2}a^{4}\left(\frac{m_{p}}{M_{\star}}\right)^{2}\left(\frac{h}{a}\right)^{-2}\label{eq:migtorque}
\end{equation}
Where $\Sigma_{g}=2\rho_{g}h$ is the gas surface density. Both torques
depend on the planetesimal mass and disk parameters on the same way.
The ratio between both torques is $T_{GDF}/T_{migration}\sim(h/a)^{-1}$.
The factor $h/a$ comes from the fact that the \emph{differential
torque} scales with $h/a$ \citep{1986Icar...67..164W}.
This is because there is an intrinsic asymmetry between inner and
outer torques. In some sense, this is analogous to tidal forces that
rise due to asymmetry of the gravitational forces induced on a solid
body. One can better see the relation between these approaches by
estimating the \emph{one sided} torque using the
GDF formulation, where the relevant relative velocity is the sound
speed and the relevant formula is the supersonic one since effective
Lindblad resonances resides at locations where the relative velocity
between the flow and the planetesimal is equal to the sound speed
\citep{1993ApJ...419..155A}. 

Finally, in our study we consider disk-planetesimal interactions
on eccentric orbits. Note that the effect of eccentricity on planet
migration was discussed in several studies such as \citealp{2000MNRAS.315..823P}.
They find that for eccentricity larger than $\sim~1.1H_{0}$ the torque
reverses. It has been confirmed by \citet{2007AA...473..329C} and \citet{2010AA...523A..30B},
but the embedded planet is always migrating inward. The reason is
that the torque changes the eccentricity, while the migration rate
is calculated from the power, which is always negative for isothermal
disks. In our models we always see inward migration.

\subsection{Implications of gas dynamical friction for planet formation and the
evolution of protoplanetary disks}

\textbf{Planetesimal disk evolution:} When considering the evolution
of large and small planetesimals in a disk, the evolution of the velocity
dispersion is governed by viscous heating of the large bodies by themselves
and their cooling by dynamical friction, whereas small planetesimals
mostly heat through dynamical friction by the large bodies \citep{2002Natur.420..643G}.
Introducing gas to the system gives rise to additional cooling channels
both for large and small bodies. For large oligarchs, GDF keeps the
random velocities low, which prevents oligarch collisions. For small
bodies, aerodynamic gas drag is the main process which dominates their
cooling; such cooling is more than two order of magnitudes faster
than cooling of small planetesimals through inelastic collisions \citep{2002Natur.420..643G}
considered before. Moreover, the drag force increases with increasing
velocity, so any random velocity raised by viscous stirring will be
suppressed by gas drag. GDF therefore keeps planetesimal disks thin
and cool much longer than otherwise thought, and can not be neglected.

For the upper tail of large ptotoplanets, the mass is close to the
isolation mass $M_{iso}\sim0.07M_{\oplus}(a/AU)^{3}(\Sigma_{g}/10g\cdot cm^{-3})^{3/2}$
where the protoplanet has cleared its feeding zone \citep{1996Icar..124...62P}.
A large protoplanet under GDF force could migrate to a new environment
where additional-planetesimal swarms are available for accretion.
Hence the protoplanets do not grow in isolation and its growth is
not limited by its local feeding zone.

Higher relative velocities can also affect the embryonic migration
rate and eccentricity and inclination damping. While initially eccentric
orbit is rapidly circularized in disks around single stars, planetesimals
embedded in disks of binary stellar systems can have persistent higher
relative velocities. The origin of high relative velocity can be either
eccentric co-planar disk, or inclined orbit due to misaligned stellar
companion \citep{2015ApJ...798...69R}.

\textbf{Super-Earth / hot Neptune formation:}

The observed distributions of planets in short periods suggest that
lower mass super-Earths / hot Neptunes are more common than gas giants,
compared with simple expectations from population synthesis models
\citep{2008ApJ...685..584I,2012ApJ...751..158H}. In order to reproduce
the observed mass distribution an over-abundance of rocky material
is required in the inner ($<1$AU) region. This can be achieved either
by radial drift of rocky material in the form of dust or small planetesimals
\citep{2012ApJ...751..158H}, or by enhanced primordial minimal mass
extrasolar nebula (MMEN) surface density profile \citep{2013MNRAS.431.3444C}.
Other suggestions involve type I migration of planets formed in the
outer region into the inner region, which requires additional physical
processes to concede with observations (i.e. eccentricity damping,
tidal friction, pressure maxima trapping, \citealp{2014arXiv1412.4440I}).
Both suggestions suffer from theoretical and observational challenges:
migration of small planetesimals through aerodynamic gas-drag, might
be too fast and lead to the accretion of the solid material by the
star, while enhanced primordial MMEN is unlikely and might be unstable
\citep{2014arXiv1412.4440I}. Migrating embryos due to GDF could provide
an additional channel for supply of solid material into the inner
region in the form of intermediate size planetesimal/planetary-embryos.
GDF embryonic migration naturally introduces eccentricity damping,
and thus alleviates the need for external mechanisms for eccentricity
damping. Moreover, slightly eccentric orbit results in a more efficient
migration, hence the migration starts at the scale of intermediate
size planetesimals. In addition, for a configuration of eccentric
disk in binary star system, higher relative velocity are predominant
for the entire disk lifetime, and the embryonic migration rate might
be faster and cause even more favorable conditions for generating
super-Earths in circumbinary disks.

These embryos could therefore assist in providing the required over-abundance
of rocky material for in-situ formation of hot Neptunes. In addition,
the time-scales involved are comparable with disk lifetime, hence
the migration is not too efficient as to lead to material accretion
into the star, but its time-scales are sufficiently short as to supply
the material during the lifetime of the gaseous disk.

\section{Summary }

\label{sec:Summary}

In this study we considered the effect of GDF on single IMPs. We find
that GDF is the dominant drag force affecting the evolution of planetesimals
larger than $10^{21}g$, for which aerodynamic gas drag effects are
negligible. We explored the effect of GDF in the linear regime and
in the mass range where accretion and non linear effects are negligible
(up to $\sim10^{25}g$). We estimated the typical time-scales for
the evolution of the orbital parameters of the planetesimals due to
GDF, and further studied them using detailed numerical simulations.
The main results can be summarized as follows.

Planetary embryos of mass $10^{24}-10^{25}g$ dissipate their inclinations
and circularize in less than a Myr, regardless of initial parameters.
Moreover, they migrate inward on time-scales comparable to the gaseous
disk lifetime $\sim5Myr$. Such embryonic migration may help explain
the origin of close-in super-Earth planets, that might form and grow
in the inner parts of the protoplanetary disk from such embryos \citep[see also][for related issues]{2012ApJ...751..158H}.
Smaller planetesimals ( $m\sim10^{23}g$) may dissipate low initial
eccentricities/inclinations ($e_{0}\lesssim0.3$, $I_{0}\lesssim0.1$rad).
Planetesimals in the lowest mass range $(\sim10^{21}g)$ can dissipate
low initial eccentricities $(e_{0}\lesssim0.1)$, and slightly damp
small inclinations (comparable to the disk scale height). The efficient
damping of inclination and eccentricity reduces the planetesimal random
velocities and assist in keeping the planetesimal disk flatter and
more circular, exchanging the planetesimals kinetic energy into the
gaseous disk. Such evolution could therefore have implication not
only for the planetesimal disk structure but also for the long term
collisional evolution and planetary growth in the disk. We conclude
that GDF can play an important role in the evolution of planetesimal
disk, and should be accounted for in the study of the early stages
of planet formation.

\section*{Acknowledgements}
We thank Wilhelm Kley for stimulating discussions and the referee, Takayuki Muto, for helpful comments that lead to improvement of the manuscript. HBP acknowledges support from Israel-US bi-national science foundation, BSF grant number 2012384, European union career integration grant "GRAND", the the Minerva center for life under extreme planetary conditions and the Israel science foundation excellence center  I-CORE grant 1829.

\bibliographystyle{apj} %\bibliography{refsingle}

\expandafter\ifx\csname natexlab\endcsname\relax\global\long\def\natexlab#1{#1}
\fi

\appendix
%dummy comment inserted by tex2lyx to ensure that this paragraph is not empty
{}

\section{A. TYPICAL ECCENTRICITY IN THE TRANS-SONIC REGIME}

\label{sec:Typical-eccentricity-for}

We wish to estimate the critical eccentricity $e_{c}$ for trans-sonic
regime. The calculation is similar to \citet{2011ApJ...737...37M},
where they neglect the pressure gradients of the gas and assume $\eta=0$.
The relative velocities in polar coordinates are 
\begin{eqnarray}
v_{rel,r} & = & v_{r,gas}-v_{r}=-\frac{\Omega a}{\sqrt{1-e^{2}}}e\sin f\label{eq:relr}\\
v_{rel,\phi} & = & v_{\phi,gas}-v_{\phi}=\sqrt{1-\eta}\Omega a-\frac{\Omega a}{\sqrt{1-e^{2}}}(1+e\cos f)=\Omega a\left[\sqrt{1-\eta}-\frac{(1+e\cos f)}{\sqrt{1-e^{2}}}\right]\label{eq:relphi}
\end{eqnarray}
The implicit assumption in obtaining (\ref{eq:relr}) and (\ref{eq:relphi})
is that the large planetesimals are unaffected by GDF on a scale of
orbital period. More formally, the \emph{stopping} \emph{time} is
large compared to the orbital period. Indeed, (\ref{eq:relr}) and
(\ref{eq:relphi}) at zero eccentricity are obtained by letting $t_{s}\Omega\gg1$
in Eqs. (16) and (17) in \citet{per+11} and taking the first order.

The total relative velocity is 
\begin{eqnarray}
v_{rel}^{2} & = & v_{rel,r}^{2}+v_{rel,\varphi}^{2}=\frac{\widetilde{a}(e,\eta)+\widetilde{b}(e,\eta)\cos f}{1-e^{2}}v_{K}^{2}\label{eq:vtot}
\end{eqnarray}
Where $\widetilde{a}(e,\eta)\equiv2-\eta+{\eta}e^{2}-2(1-\eta)^{1/2}(1-e^2)^{1/2}$
and $\widetilde{b}(e,\eta)\equiv2e-2e\boldsymbol(1-\eta)^{1/2}({1-e^{2}})^{1/2}$.  Note that $\widetilde{a}$ is a dimensionless quantity, not to be confused with the semi-major axis $a$. In order to acquire
the average relative velocity, we need to average Eq. (\ref{eq:vtot})
on one orbital period: 
\[
\langle v_{rel}^{2}(f(t),e,\eta)\rangle\equiv\frac{\Omega}{2\pi}\intop_{0}^{1/\Omega}v_{rel}^{2}(f,e,\eta)dt=\frac{\Omega}{2\pi}\intop_{0}^{2\pi}v_{rel}^{2}(f,e,\eta)\frac{dt}{df}df
\]
Where $dt/df=a(1-e^{2})^{3/2}/(v_{K}(1+e\cos f)^{2})$. The average
velocity is 
\begin{eqnarray*}
\langle v_{rel}^{2}(e)\rangle & = & \frac{\sqrt{1-e^{2}}}{2\pi}v_{K}^{2}\left[\widetilde{a}\intop_{0}^{2\pi}\frac{df}{(1+e\cos f)^{2}}+\widetilde{b}\intop_{0}^{2\pi}\frac{\cos f}{(1+e\cos f)^{2}}df\right]
\end{eqnarray*}
Which can be rewritten as 
\begin{equation}
\langle v_{rel}^{2}(e)\rangle=v_{K}^{2}(1-e^{2})^{3/2}(2-\eta - 2\sqrt{1-\eta}\sqrt{1-e^{2}})\frac{1}{2\pi}\intop_{0}^{2\pi}\frac{df}{(1+e\cos f)^{2}}\label{eq:vtotavg}
\end{equation}
Expanding in powers of $e$ we get 
\begin{equation}
\langle v_{rel}(e)\rangle=v_{K}\left[(1-\sqrt{1-\eta})^{2}+\sqrt{1-\eta} e^{2}+O(e^{4})\right]^{1/2}\label{eq:vrel}
\end{equation}
For very low eccentricities $e\ll\varepsilon$, we get $\langle v_{rel}(e)\rangle=\varepsilon v_{K}$.
For higher eccentricities $e\gg\varepsilon$ we get $v_{rel}\approx ev_{K}$.
The critical eccentricity $e_{c}$ for which the flow is trans-sonic
is $e_{c}\approx2H_{0}=0.044$, consistent with \citet{2011ApJ...737...37M}.

\section{B. NUMERICAL SET UP}

\label{sec:Numerical set up}

At each step we first evaluate $c_{s}=0.022\cdot a^{2/7}$ and $\rho_{g}=1.7\cdot10^{-3}$
at $1$AU in simulation units. For most runs we model the gas density
with a power-law slope of $\rho_{g}\sim a^{-16/7}$ . Steeper slopes
are considered in Section \ref{sub:Disk-Density-Profile}. Next, we
evaluate $\eta=19/7(c_{s}/v_{K})^{2}$
and the Cartesian representation of the gas velocity $v_{x,gas}=-v_{r,gas}\sin f=-(1-\eta)^{1/2} v_{K}y/a$
as well as $v_{y,gas}=v_{r,gas}\cos f=(1-\eta)^{1/2} v_{k}x/a$, where $f$
is the true anomaly. Finally, we evaluate $\boldsymbol{v}_{rel}=\boldsymbol{v}_{p}-\boldsymbol{v}_{gas}$
, the scalar $v_{rel}=|\boldsymbol{v}_{rel}|$, and compute the Mach
number by $\mathcal{M}=v_{rel}/c_{s}.$

The GDF coefficient $C_{GDF}$ which is calculated for every case
is $4\pi\rho_{g}\mathcal{I}(\mathcal{M})$ where $\mathcal{I}(\mathcal{M})$
is $0.5\ln\left((1+\mathcal{M})/(1-\mathcal{M})\right)-\mathcal{M})$
in the subsonic regime, and $\ln\Lambda+0.5\ln(1-\mathcal{M}^{-2})$
in the supersonic regime. The effective Coulomb logarithm is $\ln\Lambda=\ln c_{s}t/R_{min}$.
Taking $t=1yr$ and $R_{min}$ to be the planetesimal size, we get
a Coulomb logarithm of the order of $10$. In a small interval near
$\mathcal{M\sim}1$ we set $\mathcal{I}(\mathcal{M})=10$ to avoid
singularity and make GDF function continuous.

The GDF coefficient is inserted into the acceleration per unit mass
$\boldsymbol{da}=C_{GDF}\boldsymbol{v}_{rel}/v_{rel}^{3}$ and jerk
per unit mass 
\begin{eqnarray*}
\boldsymbol{dj} & = & \frac{d}{dt}\left(C_{GDF}\cdot\frac{\boldsymbol{v}_{rel}}{v_{rel}^{3}}\right)\\
 & = & \frac{4\pi\rho_{g}}{v_{rel}^{3}}\left[\frac{(\boldsymbol{v}_{rel}\cdot\boldsymbol{da})}{v_{rel}}\boldsymbol{v}_{rel}\left(\frac{1}{c_{s}}\frac{d\mathcal{I}(\mathcal{M})}{d\mathcal{M}}-\frac{3}{v_{rel}}\mathcal{I}(\mathcal{M})\right)+\mathcal{I}(\mathcal{M})\boldsymbol{da}\right]
\end{eqnarray*}
where $d\mathcal{I}(\mathcal{M})/d\mathcal{M}$ is $\mathcal{M}^{2}/(1-\mathcal{M}^{2})$
in the subsonic regime and $-2/(\mathcal{M}(1-\mathcal{M}^{2}))$
in the supersonic regime.

The total acceleration and jerk are $\boldsymbol{a}=-m_{p}\boldsymbol{da}$
and $\boldsymbol{j}=-m_{p}\boldsymbol{dj}$ respectively. 
\end{document}